\documentclass[journal]{IEEEtran}
\usepackage[cmex10]{amsmath}
\usepackage{eqparbox}
\usepackage[usenames, dvipsnames]{color}
\usepackage{mathtools,amssymb,bm,mathabx}
\usepackage{amstext}
\usepackage{amssymb}
\usepackage{graphicx}
\usepackage{color}
\usepackage{booktabs}
\usepackage{longtable}
\usepackage{multicol}
\usepackage{multirow}
\usepackage{amsfonts}
\usepackage{dsfont}
\usepackage{array}
\usepackage[ruled]{algorithm}
\usepackage{algorithmic}
\usepackage{cases}
\usepackage{pgfplots}
\usepackage{accents}
\usepackage{caption}
\usepackage[footnotesize]{subfigure}
\usepackage{amsthm,xparse}
\DeclareCaptionLabelSeparator{periodspace}{.\quad}
\usepackage{float}
\usepackage{cite}
\usepackage [autostyle, english = american]{csquotes}
\usepackage{hyperref}
\theoremstyle{remark}

\newcommand\ASTART{\bigskip\noindent\begin{minipage}[b]{0.5\linewidth}}
	
	\newcommand\AENDSKIP{\end{minipage}\bigskip}
\newcommand\AEND{\end{minipage}}
\ifCLASSOPTIONcaptionsoff
\usepackage[nomarkers]{endfloat}
\let\MYoriglatexcaption\caption
\renewcommand{\caption}[2][\relax]{\MYoriglatexcaption[#2]{#2}}
\fi
\theoremstyle{plain}
\newtheorem{thm}{\textbf{Theorem}}
\newtheorem{lem}{\textbf{Lemma}}

\theoremstyle{definition}
\newtheorem{defn}{\textbf{Definition}}

\theoremstyle{remark}
\newtheorem{rem}{\bf Remark}

\newcommand*{\rom}[1]{\expandafter\@slowromancap\romannumeral #1@}

\newcommand{\RN}[1]{%
\textup{\uppercase\expandafter{\romannumeral#1}}%
}

\usepackage{standalone}
\graphicspath{{./Figures/}} 
\begin{document}
%
\title{A Greedy Algorithm for Matrix Recovery with Subspace Prior Information}
\author{Hamideh.S~Fazael~Ardakani, Sajad~Daei, Farzan~Haddadi%
	\thanks{H S. Fazael Ardakani and S. Daei and F. Haddadi are with the School of Electrical Engineering, Iran University of Science \& Technology.}}
\maketitle

\begin{abstract}
	Matrix recovery is the problem of recovering a low-rank matrix from a few linear measurements. Recently, this problem has gained a lot of attention as it is employed in many applications such as Netflix prize problem, seismic data interpolation and collaborative filtering. In these applications, one might access to additional prior information about the column and row spaces of the matrix. These extra information can potentially enhance the matrix recovery performance. In this paper, we propose an efficient greedy algorithm that exploits prior information in the recovery procedure. The performance of the proposed algorithm is measured in terms of the rank restricted isometry property (R-RIP). Our proposed algorithm with prior subspace information converges under a more milder condition on the R-RIP in compared with the case that we do not use prior information. Additionally, our algorithm performs much better than nuclear norm minimization in terms of both computational complexity and success rate. 
\end{abstract}


%
\IEEEpeerreviewmaketitle

\section{Introduction}\label{introduction}
The problem of recovering a low-rank matrix from undersampled measurements arises in many applications such as MRI\cite{haldar2010spatiotemporal},\cite{zhao2010low}, collaborative filtering \cite{srebro2010collaborative}, Netflix \cite{bennett2007netflix}, exploration seismology \cite{aravkin2014fast}, and quantum state tomography\cite{gross2010quantum}. 

An idealistic approach to solve this problem is to consider the problem\footnote{In this work, we consider only square matrices. However, the extension to non-square matrices is straightforward.}
\begin{align}\label{p.1}
&\min_{\bm{Z} \in \mathbb{R}^{n\times n}}~{\rm{rank}}(\bm{Z}) \nonumber \\
&\mathrm{s.t.}~\bm{y} = \mathcal{A}(\bm{Z}),
\end{align}
which is known to be NP-hard in general \cite{recht2010guaranteed}.
Here $ \mathcal{A}:\mathbb{R}^{n\times n} \rightarrow \mathbb{R}^{p} $ is a linear operator, and $ \bm{y} \in \mathbb{R}^{p}$ for $ p $ measurement. Then problem \eqref{p.1} can be relaxed and converted to the following tractable convex problem:
\begin{align}\label{p.2}
&\min_{\bm{Z} \in \mathbb{R}^{n\times n}}~\|\bm{Z}\|_{*} \nonumber \\
&\mathrm{s.t.}~\bm{y} = \mathcal{A}(\bm{Z}),
\end{align}
where $ \|\cdot\|_{*}$ sums the singular values of a matrix and is called nuclear norm. There are many algorithms based on SVD, truncated SVD, and greedy methods for matrix recovery and weighted matrix recovery (see Section \ref{Related Works} for more explanations); however these algorithms are not capable to exploit prior subspace information. In this paper, we propose an algorithm that uses prior information for low-rank matrix recovery based on CoSaMP \footnote{Compressive Sampling Matching Pursuit}\cite{needell2009cosamp}. Before introducing our algorithm for matrix recovery, it is necessary to know how to incorporate prior information into matrix recovery problem. Let $\bm{\mathcal{U}}_r$ and $\bm{\mathcal{V}}_r$ be the column and row space of the ground-truth matrix $\bm{X}$ with rank $r$, respectively. Suppose that we are given two subspaces $\widetilde{\bm{\mathcal{U}}}_r$ and $\widetilde{\bm{\mathcal{V}}}_r$ that form the principal angles\footnote{For the definition of principal angles between subspaces, see for example\cite[Section II]{daei2018optimal}.} 
\begin{align*}
\mathbf{\bm{\theta}}_{u}= \angle[\bm{\mathcal{U}_r},\widetilde{\bm{\mathcal{U}}}_r],~ \mathbf{\bm{\theta}}_{v}=\angle[\bm{\mathcal{V}}_r,\widetilde{\bm{\mathcal{V}}}_r],
\end{align*}
with $\bm{\mathcal{U}}_r$ and $\bm{\mathcal{V}}_r$, respectively. We also define the subspaces ${\widetilde{\bm{\mathcal{U}}}_{r}}^{\perp}$ and ${\widetilde{\bm{\mathcal{V}}}_{r}}^{\perp}$ as the orthogonal complements of ${\widetilde{\bm{\mathcal{U}}}_{r}}$ and ${\widetilde{\bm{\mathcal{V}}}_{r}}$, respectively.
Then, the following problem is proposed for capturing both low-rankness and subspace prior information:
\begin{align}\label{p.4}
&\min_{\bm{Z} \in \mathbb{R}^{n\times n}} ~{\rm rank}(\bm{Q}_{\widetilde{\bm{\mathcal{U}}}_{r}} \bm{Z} \bm{Q}_{\widetilde{\bm{\mathcal{V}}}_{r}}) \nonumber \\
&\mathrm{s.t.}~\bm{y} = \mathcal{A}(\bm{Z}),
\end{align}
where
\begin{align}\label{3}
&\bm{Q}_{\widetilde{\bm{\mathcal{U}}}_{r}} := \widetilde{\bm{U}}_{r}\bm{W}_{1}\widetilde{\bm{U}}_{r}^{\rm{H}} +\widetilde{\bm{U}}_{r}^{\perp}\bm{W}_{2}\widetilde{\bm{U}}_{r}^{\perp H} \nonumber \\
&\bm{Q}_{\widetilde{\bm{\mathcal{V}}}_{r}} := \widetilde{\bm{V}}_{r}\bm{W}_{3}\widetilde{\bm{V}}_{r}^{\rm{H}} + \widetilde{\bm{V}}_{r}^{\perp}\bm{W}_{4}\widetilde{\bm{V}}_{r}^{\perp H},
\end{align}
and $ \bm{W}_{i}, i = 1,..., 4 $ are diagonal matrices. Here, $\widetilde{\bm{U}}_{r}$ and $\widetilde{\bm{U}}_{r}^{\perp}$ are some bases of the subspaces  ${\widetilde{\bm{\mathcal{U}}}_{r}}$ and  ${\widetilde{\bm{\mathcal{U}}}_{r}}^{\perp}$, respectively (the same argument holds for $\widetilde{\bm{V}}_{r}$ and $\widetilde{\bm{V}}_{r}^{\perp}$). The diagonal entries of $ \bm{W}_{i} $s specify how much the corresponding principal angles shall be penalized in the minimization problem. For instance, if $\widetilde{\bm{\mathcal{U}}}_{r}$ is close to $\bm{\mathcal{U}}_r$ and far from the $\bm{\mathcal{U}}_r^{\perp}$, then the values on the diagonal of $\bm{W}_1$ are small while the values on the diagonal of $\bm{W}_2$ shall be large.

\subsection{Contributions}
To highlight our contributions, we list them below:
\begin{enumerate}
	\item \textit{Proposing a new optimization problem for matrix recovery and completion.}
	As mentioned in \eqref{p.4}, we design a novel optimization problem that encourages both rank and subspace information. The problem uses the principal angles between a given subspace and the matrix subspace.
	\item \textit{Proposing a greedy-based algorithm for matrix recovery and completion}. 
	We propose new and efficient greedy-based algorithms named rank minimization with subspace prior information (RMSPI) and generalized RMSPI (GRMSPI) to solve \eqref{p.4}. While RMSPI penalizes the principal angles with a single weight, GRMSPI is designed for a multi-weight scenario where each principal angle is penalized separately.
	\item \textit{Convergence guaranty.} We prove convergence guarantee results for RMSPI and GRMSPI. The convergence rate of our proposed algorithm is superior to that of \cite{lee2010admira}.
	\item We present a performance guarantee for RMSPI and GRMSPI in terms of rank-restricted isometry property (R-RIP) given in Section \ref{section.mainresult}. Simulation results show that even when the measurement operator does not satisfy the R-RIP constraint (for example in matrix completion problem), our proposed algorithms are still capable of recovering the interested matrix exactly.
	\item We examine our algorithms in presence of noisy measurements and numerically observe that it is robust to measurement noise.
\end{enumerate}
\subsection{Related Works and Key Differences}\label{Related Works}
Recht et al. in \cite{recht2010guaranteed} convert problem \eqref{p.1} to the relaxed form in \eqref{p.2}. 
Following \cite{recht2010guaranteed}, Cai et al. in \cite{cai2010singular} develop an algorithm for solving the problem
\begin{align}\label{p.5}
\min _{\bm{Z} \in \mathbb{R}^{n\times n}} \|\mathcal{A}(\bm{Z}) - 
\bm{y}\|_{2}^{2} + \lambda\|\bm{Z}\|_{*}
\end{align}
which is a regularized version of \eqref{p.2}. Specifically, they provide an iterative algorithm based on a soft singular value thresholding (SVT). Besides the fact that the algorithm is designed for the noiseless case, the computation cost is rather low, yet it lacks any convergence rate analysis.

Ji et al. in \cite{ji2009accelerated}, Liu et al. in \cite{liu2012implementable}, and Toh et al. in\cite{toh2010accelerated} independently provide algorithms based on gradient method to solve the problem \eqref{p.5}. However, the purpose of these papers is to improve SVT and to reduce the number of required iterations for matrix recovery.

Mazumder et al. in \cite{mazumder2010spectral} propose a convex algorithm for minimizing the error in each iteration under the condition that the nuclear norm is bounded. This algorithm requires taking SVD in each step which is costly.

Ma et al. in \cite{ma2011fixed} propose an iterative algorithm for minimizing nuclear norm by using the Bregman divergence and fixed point. This algorithm is very fast and powerful, yet suffers lack of convergence rate analysis. Also, it is only useful in the absence of noise.

In \cite{mishra2013low}, \cite{wen2010low}, and \cite{recht2013parallel}, the authors propose methods based on approximating the nuclear norm by using its variational features.

The above-mentioned methods need calculating SVD which is expensive in general. Henceforth, in \cite{wang2015orthogonal} and \cite{lee2010admira}, the authors propose algorithms based on greedy methods. More explicitly, they extend the well-studied methods OMP\footnote{Orthogonal Matching Pursuit}\cite{tropp2007signal} and CoSaMP \cite{needell2009cosamp} to the matrix case. These methods are observed to be more efficient than relaxation-based algorithm (e.g. nuclear norm minimization).

Despite the effectiveness in matrix recovery problem, only few works consider prior information \cite{jain2013provable}, \cite{xu2013speedup}, \cite{eftekhari2018weighted}. Specifically, the common feature of these algorithms is to use prior information in such a way that the number of required measurements is minimized. 

The authors in \cite{jain2013provable}, \cite{xu2013speedup}, employ side information to enhance the performance of nuclear norm minimization. Their side information was to completely know a few directions of the matrix subspace and differs from knowing the principal angles that we consider.

Aravkin et al. in \cite{aravkin2014fast} and Eftekhari et al. in \cite{eftekhari2018weighted} incorporate prior knowledge about the matrix column and row spaces into the recovery procedure. They consider the maximum principal angle between a given (e.g. $\widetilde{\bm{\mathcal{U}}}_r$) and the ground-truth subspace (e.g. $\bm{\mathcal{U}}_r$). They show that as long as the given subspace is close to the interested one, the required number of measurements decreases compared to the regular nuclear norm minimization. Our model in \eqref{p.4} differs from the ones in \cite{aravkin2014fast} and \cite{eftekhari2018weighted} in that we penalize the given subspace with multiple weights instead of a single weight. Also, our model outperforms theirs when $\widetilde{\bm{\mathcal{U}}}_r$ is either far or close to $\bm{\mathcal{U}}_r$. Besides the generality of our model, RMSPI and GRMSPI are considerably superior to the ones in \cite{aravkin2014fast,eftekhari2018weighted} in terms of computational complexity (see Section \ref{section.simulation}).

The authors in \cite{mohan2010reweighted}, propose an alternative for rank minimization:  
\begin{align}
&\min_{\bm{Z} \in \mathbb{R}^{n\times n}} \log(\det(\bm{Z})) \nonumber \\
& \bm{Z} \in C.
\end{align}
where $C$ is the feasible set. 
In each iteration (e.g. $k~$th) of the algorithm solves
\begin{align}
&\min_{\bm{Z} \in \mathbb{R}^{n\times n}}~\|\bm{W}_{1}^{k}\bm{Z}\bm{W}_{2}^{k}\| \nonumber \\
& \bm{Z} \in C,
\end{align}
where $\bm{W}_{1}^{k}$ and $\bm{W}_{2}^{k}$ are some weighting matrices.

Finally, \cite{rao2015collaborative} solves
\begin{align}
\min_{\bm{Z} \in \mathbb{R}^{n\times n}} \dfrac{1}{m}\|\mathcal{A}(\bm{Z})-\bm{y}\|_{2} + \lambda_{N}\|\bm{A}\bm{Z}\bm{B}\|_{*},
\end{align}
in which $ \bm{A} $ and $ \bm{B} $ are some certain invertible matrices related to the interested subspace. This problem is also similar to \cite{ji2009accelerated}.

\subsection{Outline and Notations}

The paper is organized as follows: problem formulation and the proposed algorithm are given in Section \ref{section.explanation of algorithm}. In Section \ref{section.mainresult}, we explain the main result regarding the convergence rate of our algorithm. We compare our proposed methods with the state of the art algorithms in Section \ref{section.simulation}.

Throughout the paper, scalars are denoted by lowercase letters, vectors by lowercase boldface letters, and matrices by uppercase letters. The $i$th element of the vector $\bm{x}$ is shown by $x(i)$ or $x_{i}$. The operators $\text{Tr}(\cdot)$ and $(\cdot)^{\rm{H}}$ are used to denote the trace and Hermitian of a matrix, respectively. $ (\cdot)^{\dagger} $ represents the pseudo-inverse operator. The Frobenius inner product is denoted by $\langle \bm{A}, \bm{B} \rangle_{F}= \text{Tr}(\bm{A}\bm{B}^{\rm{H}})$. The orthogonal projection matrices onto the subspaces $\bm{\mathcal{U}}$ and $\bm{\mathcal{U}}^{\perp}$ are shown by $$\bm{P}_{\bm{\mathcal{U}}} := \bm{{U}}\bm{{U}}^{\rm{H}},$$ and $$\bm{P}_{\bm{\mathcal{U}}^{\perp}} := \bm{I} - \bm{P}_{\bm{\mathcal{U}}},$$
where $\bm{U}$ is a basis fo r the subspace $\bm{\mathcal{U}}$ and $\bm{I}$ is the identity matrix. Also define the support of $\bm{X}$ by the linear subspace
$$ \bm{T} = \{\bm{Z}\in\mathbb{R}^{ m\times n} : \bm{Z}=\bm{P}_{\bm{\mathcal{U}}}\bm{Z}\bm{P}_{\bm{\mathcal{V}}}+\bm{P}_{\bm{\mathcal{U}}}\bm{Z}\bm{P}_{\bm{\mathcal{V}}^\perp} +\bm{P}_{\bm{\mathcal{U}}^\perp}\bm{Z}\bm{P}_{\bm{\mathcal{V}}^\perp}\}$$
$$:={\rm supp(\bm{X})}$$
We define the linear operator $\mathcal{A}:\mathbb{R}^{m \times n} \rightarrow \mathbb{R}^{p}$ as $$\mathcal{A}\bm{X} =[\langle \bm{X},\bm{A}_{1}\rangle _{F} ,\dots, \langle \bm{X},\bm{A}_{p}\rangle _{F}]^{\rm T} $$ for appropriate $\bm{A}_{i} \in \mathbb{R}^{m \times n }$. Also, the adjoint operator of $\mathcal{A}$ is defined as $ \mathcal{A}^{*}\bm{y} = \sum_{i=1}^{p}y_{i}\bm{A}_{i}$. $ \mathcal{I} $ is identity linear operator i.e. $ \mathcal{I}^{*}\mathcal{I}\bm{X}= \bm{X}$.

\section{Algorithm Formulation}\label{section.explanation of algorithm} 
Let $\bm{X} \in \mathbb{R}^{n \times n }$ be a matrix with singular value decomposition (SVD) $\bm{X} =\bm{ U}_{n\times n}\bm{ \Sigma}_{n\times n} \bm{V}_{n\times n}^{\rm{H}}$. Here, $\bm{U}:=[\bm{u}_1,..., \bm{u}_n]$ and $\bm{V}:=[\bm{v}_1, ..., \bm{v}_n]$ are orthonormal bases. For an integer $r\le n$, we might have a rank $r$ truncation of $\bm{X}$ as $\bm{X}_r=\bm{U}_{r}\bm{\Sigma}_{r}\bm{V}_{r}^{\rm{H}}$ where $\bm{U}_{r}$ and $\bm{V}_r$ are obtained by retaining $r$ columns of $\bm{U}$ and $\bm{V}$ corresponding to the $r$ largest singular values i.e. $\bm{U}_{r}:=[\bm{u}_1,..., \bm{u}_r]$ and $\bm{V}_{r}:=[\bm{v}_1,..., \bm{v}_r]$. We also denote the residual by $\bm{X}_r^+:=\bm{X}-\bm{X}_r$. 
Another decomposition that we employ in this paper is the atomic decomposition. 

\begin{defn}
	Let us denote an orthonormal set of rank-1 matrices in $\mathbb{R}^{n\times n}$ by $\mathcal{O}$. We define the smallest set of rank-1 matrices in $\mathcal{O}$ that spans $\bm{X}$ as
	\begin{align}
	{\rm{atoms}}(\bm{X}) = \arg \underset{\bm{X}}{\min}\{|\Psi|:\Psi\subset \mathcal{O} , \bm{X} \in {\rm span}(\Psi)\},	
	\end{align}
	where $\Psi = \{\mathfrak{\psi} \in \mathcal{O}:\langle \mathfrak{\psi}_{j},\mathfrak{\psi}_{k}\rangle = \delta_{jk}\}$ and $ \delta $ is the indicator function and $|\cdot|$ returns the cardinality of a set.
\end{defn}

As stated earlier, we have access to some prior information about the column and row spaces of $\bm{X}$. More explicitly, we are given the subspaces $\widetilde{\bm{\mathcal{U}}}_{r}$ and $\widetilde{\bm{\mathcal{V}}}_{r}$ that form the principal angles   
\begin{align*}
\mathbf{\bm{\theta}}_{u}= \angle[\bm{\mathcal{U}}_r,\widetilde{\bm{\mathcal{U}}}_r],~ \mathbf{\bm{\theta}}_{v}=\angle[\bm{\mathcal{V}}_r,\widetilde{\bm{\mathcal{V}}}_r],
\end{align*}
with the column and row spaces of $\bm{X}$, respectively.
Diagonal values of $\bm{W}_i$s are supposed to be in the range of $[0,1]$. Notice that $\bm{\mathcal{U}}_r$ and $\widetilde{\bm{\mathcal{U}}}^{\perp}$ form $r$ distinct angles with each other. The directions corresponding to the principal angles i.e. $\bm{u}_i\bm{v}_i^{\rm H}, i=1, ..., r$ are determined based on the angles i.e. if the angles are small, the corresponding directions shall be penalized less and vice versa. However, the diagonal values of $\bm{W}_2$ and $\bm{W}_4$ corresponding to other directions (not having effective role on the principal angles) are set to $1$. We propose algorithms called RMSPI and GRMSPI to exploit this beneficial subspace information. Our proposed algorithms have very low computational cost. 
Indeed, our aim is to solve the following problem:
\begin{align}\label{p.6}
&\widehat{\bm{X}}_{\rm rec}:=\bm{Q}_{\widetilde{\bm{\mathcal{U}}}_{r}}^{-1}\mathop{\arg\min}_{\bm{Z} \in \mathbb{R}^{n \times
		n}}{\rm{rank}}(\bm{Z})\bm{Q}_{\widetilde{\bm{\mathcal{V}}}_{r}}^{-1}\nonumber \\
&\mathrm{s.t.}~\bm{y} = \mathcal{A}(\bm{Q}_{\widetilde{\bm{\mathcal{U}}}_{r}}^{-1}\bm{Z}\bm{Q}_{\widetilde{\bm{\mathcal{V}}}_{r}}^{-1}):=\mathcal{B}(\bm{Z}).
\end{align} 
This problem is mathematically equivalent to \eqref{p.4}.

In the first step of our greedy algorithms, we maximize the correlation between the residual matrix in each iteration and the atoms to update an estimate for the support of the true matrix i.e. ${\rm supp(\bm{X})}$. To do so, we maximize the norm of the projection of the residual matrix over all the subspaces i.e.
\begin{align*}
\max_{\psi \in \mathcal{O}}\{ \|\mathcal{P}_{\psi}\mathcal{A}^{*}(\bm{y}-\mathcal{A}\widehat{\bm{X}}_{\rm rec})\|_{F} \}  \\ \nonumber 
= \max_{\psi \in \mathcal{O}} |\langle (\bm{y}-\mathcal{A}\widehat{\bm{X}}_{\rm rec}),\mathcal{A}\psi\rangle|,
\end{align*}
to reach the new rank-$r$ matrix with support
\begin{align}\label{eq.support_update}
\acute{\Psi} \leftarrow \arg \underset{\Psi \subset \mathcal{O}}{\max}\{ \|\mathcal{P}_{\psi}\mathcal{A}^{*}(\bm{y}-\mathcal{A}\widehat{\bm{X}}_{\rm rec})\|_{F}  :‌ |\Psi| \leq r\}. 
\end{align}

We have provided the pseudo code of the proposed RMSPI and GRMSPI in Algorithm \ref{algorithm.proposed}. The difference between  RMSPI and GRMSPI is in choosing the matrices $\bm{Q}_{\widetilde{\bm{\mathcal{U}}}_{r}}$ and $\bm{Q}_{\widetilde{\bm{\mathcal{V}}}_{r}}$. RMSPI uses \eqref{RMSPI} for  $\bm{Q}_{\widetilde{\bm{\mathcal{U}}}_{r}}$ and $\bm{Q}_{\widetilde{\bm{\mathcal{V}}}_{r}}$ while GRMSPI uses \eqref{3}. Given the principal angles, the optimal choice of weights in $\bm{Q}_{\widetilde{\bm{\mathcal{U}}}_{r}}$ and $\bm{Q}_{\widetilde{\bm{\mathcal{V}}}_{r}}$ (either in RMSPI or GRMSPI) is challenging and beyond the scope of this paper. We, however, find the weights heuristically. In other words, we use the discussion following \eqref{3}.

The approach of our algorithm is based on the greedy method used in the vector case such as CoSaMP. At the beginning of each iteration (step 4), we obtain a set of atoms or support with size $2r$ according to \eqref{eq.support_update}. In step 6, by solving the least squares problem, a good approximation of $\bm{X}$ with rank $3r$ and known support estimate $\widetilde{\Psi}$ is obtained; for more details of least squares method see Appendix \ref{proof.least_square}. The support $\widetilde{\Psi}$ in step 5 is obtained by concatenating the support estimate in step 4 and the support estimates in the previous iteration. In the final step, since our matrix has to be of rank $ r$, we retain only the directions corresponding to the $r$ largest correlations in step 7.  In the final step, we use the least square solution to find the corresponding singular values.

\begin{algorithm}[t]
	\caption{}
	\begin{algorithmic}[1]\label{algorithm.proposed}
		\REQUIRE $\mathcal{A}:\mathbb{R}^{n\times n} \rightarrow \mathbb{R}^{p}, \bm{y}= \mathcal{A}\bm{X} \in \mathbb{R}^{p},  \widetilde{\bm{\mathcal{U}}}_{r}, \widetilde{\bm{\mathcal{V}}}_{r} $
		
		\STATE GRMSPI: Use $ \ref{3} \text{~for~} \bm{Q}_{\widetilde{\bm{\mathcal{U}}}_{r}}, \bm{Q}_{\widetilde{\bm{\mathcal{V}}}_{r}}$
		\STATE RMSPI: Use $ \ref{RMSPI} \text{~for~} \bm{Q}_{\widetilde{\bm{\mathcal{U}}}_{r}}, \bm{Q}_{\widetilde{\bm{\mathcal{V}}}_{r}}$ 
		\STATE $ \widehat{\bm{X}} \gets 0 $ 
		\STATE $\widehat{\Psi} \gets \emptyset $
		\WHILE {$ \dfrac{\|\bm{X} - \widehat{\bm{X}}_{\rm rec} \|_{F} }{\|X_{0}\|_{F}} \ge 10^{-2}  $}
		\STATE $\acute{\Psi} \leftarrow \arg \underset{\Psi \subset \mathcal{O}}{\max}\{\|\mathcal{P}_{\Psi}\bm{Q}_{\widetilde{\bm{\mathcal{U}}}_{r}}^{-1}\mathcal{A}^{*}(\bm{y}-\mathcal{A}(\widehat{\bm{X}}_{\rm rec}))\bm{Q}_{\widetilde{\bm{\mathcal{V}}}_{r}}^{-1}\|_{F} : |\Psi| \leq 2r \} $
		\STATE $ \widetilde{\Psi} \leftarrow \acute{\Psi} \cup \widehat{\Psi} $
		\STATE $ \widetilde{\bm{X}} \leftarrow \arg \underset{X}{\min}\{ \|\bm{y} -\mathcal{A}(\bm{Q}_{\widetilde{\bm{\mathcal{U}}}_{r}}^{-1}\bm{X}\bm{Q}_{\widetilde{\bm{\mathcal{V}}}_{r}}^{-1}) \|_{2}: \bm{X} \in {\rm span}(\widetilde{\Psi}) \}$
		\STATE $\widehat{\Psi} \leftarrow \arg \underset{\Psi \subset \mathcal{O}}{\max}\{ \|\mathcal{P}_{\Psi}\widetilde{\bm{X}}\|_{F} : |\Psi| \leq r \} $
		\STATE $ \widehat{\bm{X}} \leftarrow \mathcal{P}_{\widehat{\Psi}}\widetilde{\bm{X}} $
		\ENDWHILE
	\end{algorithmic} 
	Return: $ \bm{Q}_{\widetilde{\bm{\mathcal{U}}}_{r}}^{-1}\widehat{\bm{X}}\bm{Q}_{\widetilde{\bm{\mathcal{V}}}_{r}}^{-1}$ 
\end{algorithm}

\section{Main Results}\label{section.mainresult}
In this section, we investigate convergence guarantees for RMSPI and GRMSPI. We also compare our proposed algorithms with ADMiRA \cite{lee2010admira}.

Before stating our main theorem about convergence, we should emphasize that our proofs of convergence are completely different from \cite{lee2010admira} which does not use subspace prior information. Similar to section \ref{section.explanation of algorithm}, we use the equivalent problem \eqref{p.6} instead of \eqref{p.4}. First, we define the well-studied restricted isometry property (RIP) in the following.
\begin{defn}
	A linear operator $\mathcal{A}$ satisfies RIP condition with constant $\delta_{r}$ if
	\begin{align}\label{eq.RIP_A}
	(1-\delta_{r}(\mathcal{A}))\|\bm{X}\|^{2}_{F} \leq \|\mathcal{A}\bm{X}\|_{2}^{2} \leq (1+\delta_{r}(\mathcal{A}))\|\bm{X}\|^{2}_{F}
	\end{align}
	holds for every $\bm{X}$ that $ {\rm{rank}}(\bm{X}) \leq r$ \cite[Section 9.12]{foucart2017mathematical}.
\end{defn}  
In what follows, we include the relation between RIP constants of $\mathcal{A}$ and $\mathcal{B}$. 
\begin{lem}\label{lem.RIPrel}
	Let $\bm{Q}_{\widetilde{\bm{\mathcal{U}}}_{r}}$ and $\bm{Q}_{\widetilde{\bm{\mathcal{V}}}_{r}}$ be defined as \ref{RMSPI} and \ref{3} Consider the operator
	\begin{align}
	\mathcal{B}(\cdot):=\mathcal{A}(\bm{Q}_{\widetilde{\bm{\mathcal{U}}}_{r}}^{-1}\cdot\bm{Q}_{\widetilde{\bm{\mathcal{V}}}_{r}}^{-1}).	
	\end{align}
	Then, we have:
	\begin{align}
	\delta(\mathcal{A})\le \delta(\mathcal{B}).
	\end{align}
\end{lem}
Proof. See Appendix \ref{proof.lemRIP}. 
\begin{thm}\label{thm1}
	Let $ \bm{X} $ and $\widehat{\bm{X}}_{\rm rec}$ be the ground-truth matrix and the solution of \eqref{p.6}, respectively. Then, if $ \delta_{4r}(\mathcal{B})$ satisfies
	\begin{align*}
	\delta_{4r}(\mathcal{B}) < \sqrt{\dfrac{\sqrt{\frac{11}{3}}-1}{4}} \approx 0.4782,
	\end{align*}
	we have that
	\begin{align*}
	&\|(\bm{X}_{r}- \widehat{\bm{X}}_{\rm rec}^{k})\|_{F} \leq  \rho^{k}\|(\bm{X}_{r}- \bm{X}^{0})\|_{F}  \nonumber \\ 
	&+ \sqrt{\dfrac{2(1+3\delta_{4r}^{2}(\mathcal{B}))}{1-\delta_{4r}^{2}(\mathcal{B})}} \|\mathcal{P}_{\Phi\varDelta \acute{\Psi}}\mathcal{B}^{*}(\acute{\bm{e}})\|_{F} \nonumber \\
	&+ \dfrac{2}{1-\delta_{4r}(\mathcal{B})} \|\mathcal{P}_{\widetilde{\Psi}}\mathcal{B}(\acute{\bm{e}})\|_{F},
	\end{align*}
	where $0 \le \rho \le 1 $ and $\acute{\bm{e}} := \mathcal{B}(\bm{X}_{r}^{+})$.	
\end{thm}
Proof. See Appendix \ref{proof.conergancerate }.

Theorem \ref{thm1} has some consequences which are included in the following remarks.

\begin{rem}
	In ADMiRA, it is proved that if $\delta_{4r}(\mathcal{A}) \leq 0.04$ then 
	\begin{align*}
	\|(\bm{X}_{r}- \widehat{\bm{X}}_{\rm rec}^{k})\|_{F} \leq  2^{-k}\|(\bm{X}_{r})\|_{F}.
	\end{align*}
	However, by Theorem \ref{thm1}, we find that if $\delta_{4r}(\mathcal{B}) \leq 0.04$, then 
	\begin{align*}
	\|(\bm{X}_{r}- \widehat{\bm{X}}_{\rm rec}^{k})\|_{F} \leq  0.05^{k}\|(\bm{X}_{r})\|_{F},
	\end{align*}
	which verifies the superiority of  our algorithm in terms of convergence rate.
\end{rem}
\begin{rem}
	With a fixed convergence rate, the RIP constant of our operator which is equipped with subspace prior information is more conservative than that in ADMiRA. For example, to reach a convergence rate of $0.5$, $\delta_{4r}(\mathcal{A})$ in ADMiRA must be less than $0.04$ while in our algorithm, we need $\delta_{4r}(\mathcal{A})\le \delta_{4r}(\mathcal{B}) \leq 0.08966$ by Lemma \ref{lem.RIPrel}. This in turn implies that the convergence rate of our algorithm is more than that in ADMiRA. 
\end{rem}
In some applications, we do not have access to all angles between $\bm{\mathcal{U}}_r$ and $\widetilde{\bm{\mathcal{U}}}_r$, and we are only aware of the maximum angle. Thus, we have to penalize the subspace $\widetilde{\bm{\mathcal{U}}}_r$ or $\widetilde{\bm{\mathcal{U}}}_r^\perp$  with only one weight. To follow this practical model, we occupationally use 
\begin{align} \label{RMSPI}
&\bm{Q}_{\widetilde{\bm{\mathcal{U}}}_{r}} := w_{1}\bm{P}_{\widetilde{\bm{\mathcal{U}}}_{r}} + w_{2} \bm{P}_{\widetilde{\bm{\mathcal{U}}}_{r}^{\perp}} \nonumber \\
&\bm{Q}_{\widetilde{\bm{\mathcal{V}}}_{r}} := w_{3}\bm{P}_{\widetilde{\bm{\mathcal{V}}}_{r}} + w_{4} \bm{P}_{\widetilde{\bm{\mathcal{V}}}_{r}^{\perp}},
\end{align}
in Algorithm \ref{algorithm.proposed} which we call RMSPI. This model might ignore the exact penalizations. This means that all the given subspace i.e. $\widetilde{\bm{\mathcal{U}}}$ is penalized with a single weight. We investigate this simple model in Section \ref{section.simulation}.
\section{Simulation Results}\label{section.simulation}
In this section, we provide numerical experiments which compares RMSPI and GRMSPI with ADMiRA \cite{lee2010admira} in terms of the required iterations, success rate and computational complexity. Almost all of the experiments are implemented for $\bm{X}  \in \mathbb{R}^{30 \times 30}$ with rank $r = 3$, except where otherwise stated. Each experiment is repeated $50$ times over the choice of $\mathcal{A}$.   

\begin{figure*}[t]
	\centering
	\mbox{\subfigure[]{\includegraphics[width=2.34in, height=1.5in]{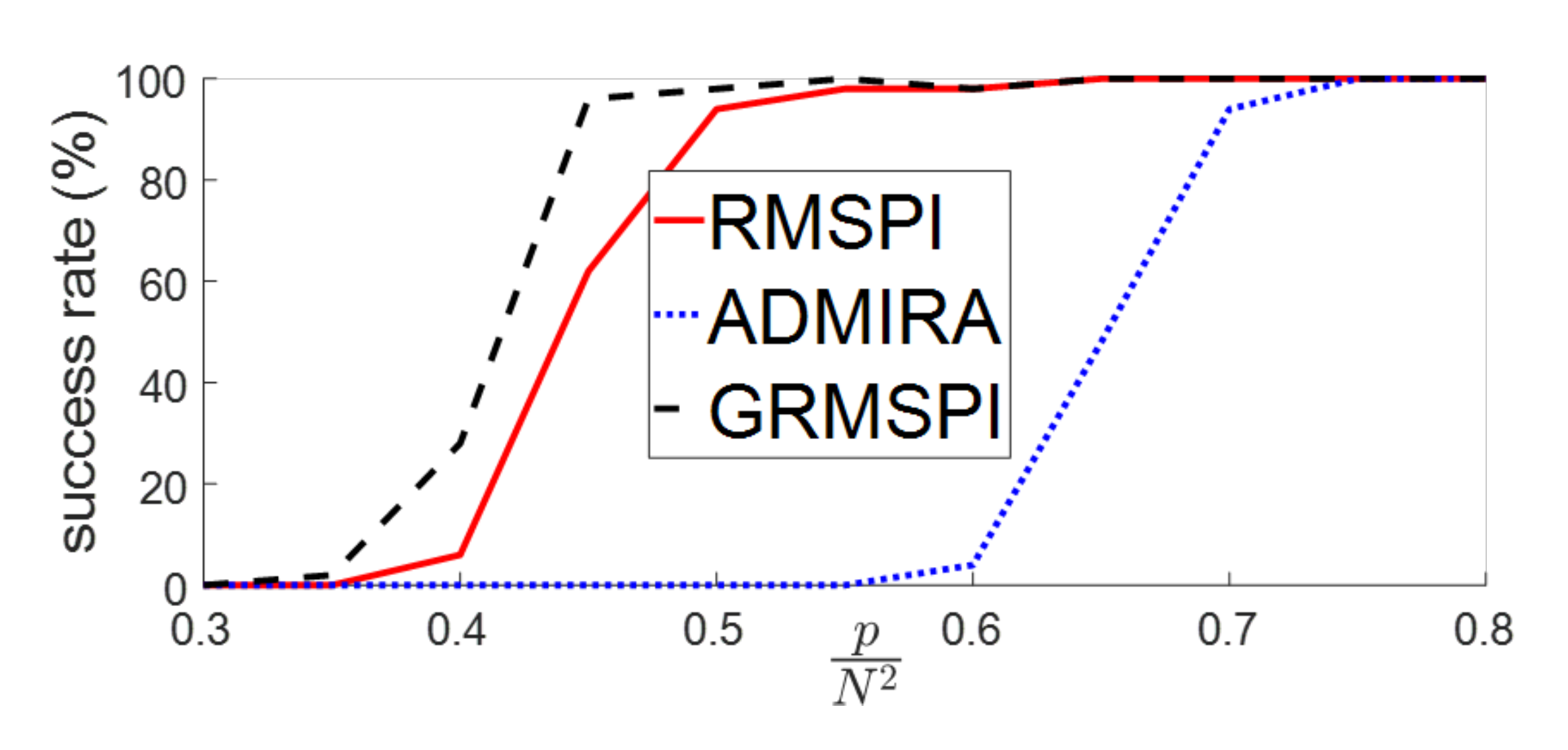}\label{fig1:a}}\quad
		\subfigure[]{\includegraphics[width=2.34in , height=1.5in]{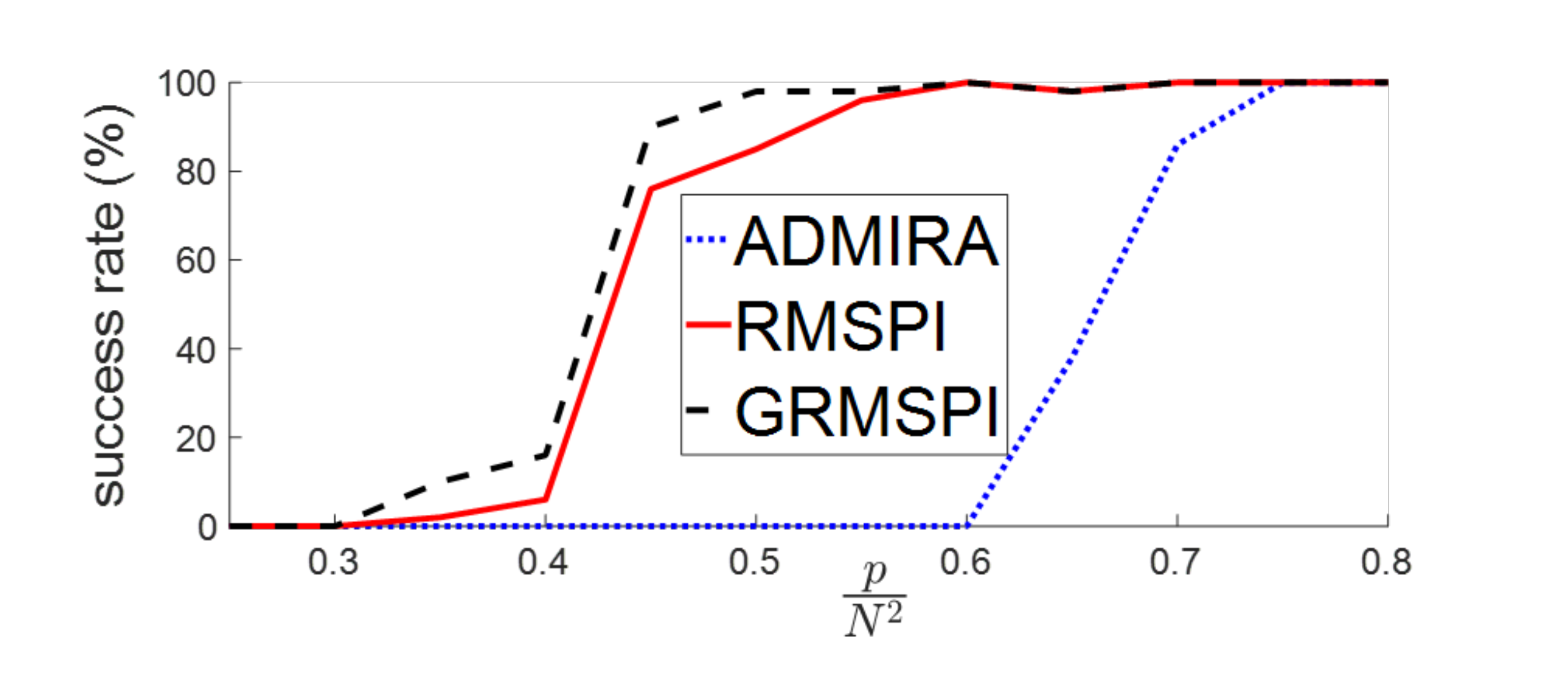}\label{fig1:b}}\quad
		\subfigure[]{\includegraphics[width=2.34in , height=1.5in]{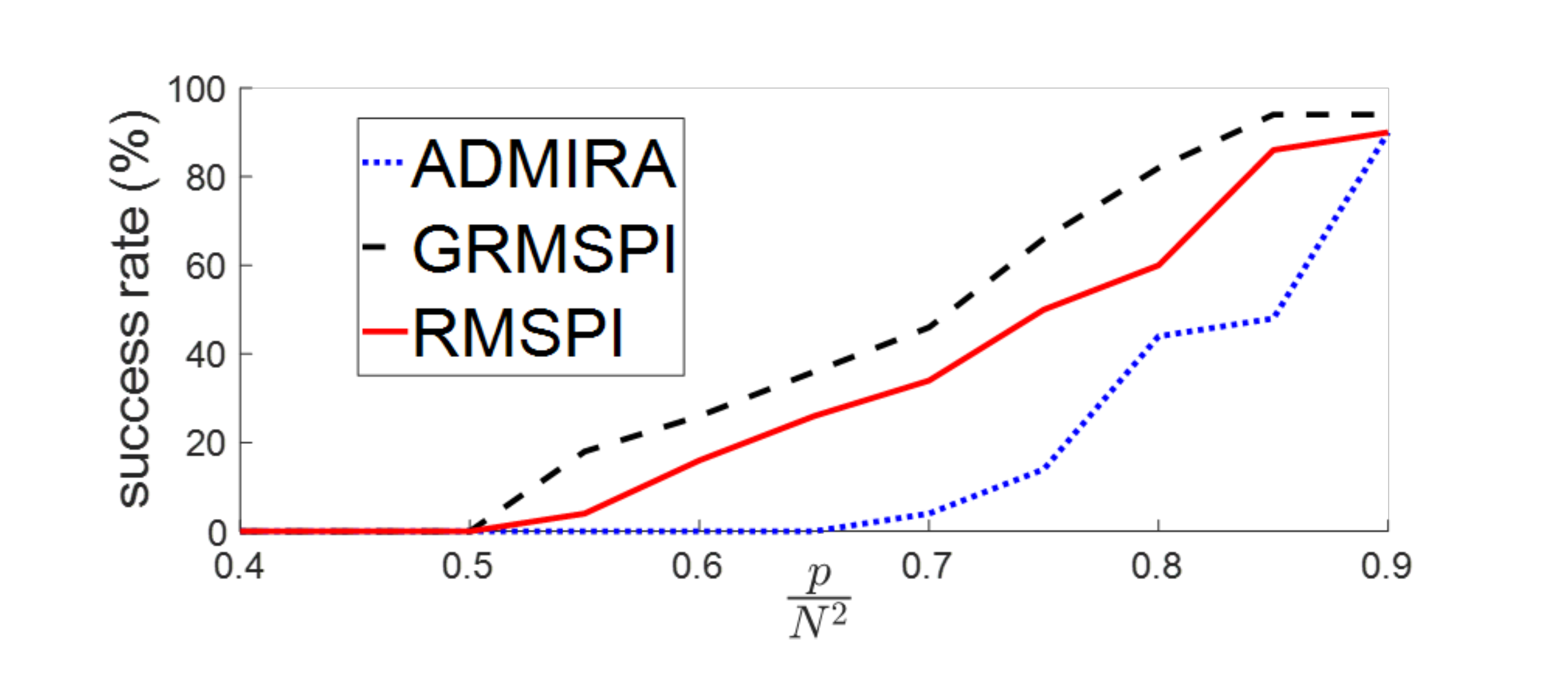}\label{fig1:c}}
	}	
	\caption{Comparison of ADMiRA with RMSPI and GRMSPI in terms of success rate for (a) noiseless matrix recovery (b) noisy matrix recovery (c) matrix completion without noise. The respective principal angles are $\bm{\theta}_{u}=[2.3307, 3.1302,3.8852]^{\rm{T}}$ and $\bm{\theta}_{v}=[2.4493, 2.9559,4.1325]^{\rm{T}}$ and the weights are $w_{1}=w_{3}=0.18$, $w_{2}=w_{4}=0.999$ ,$\bm{W}_{1}=\bm{W}_{3}={\rm  diag}[ 0.17, 0.19,0.21]$, $ \bm{W}_{2}=diag[0.99,0.98,0.97] $, $\bm{W}_{4}=diag[0.99, 0.98,0.97 ]$.}
	\label{fig1ADMIRAandprior0.05}
\end{figure*}

Figures \ref{fig1:a},\ref{fig1:b}, \ref{fig1:c} show the success rate for noisy matrix recovery, noiseless matrix recovery and noiseless matrix completion, respectively. In this experiment, we assume that $\widetilde{\bm{\mathcal{U}}}_{r}$ and $\widetilde{\bm{\mathcal{V}}}_{r}$ are close to $\bm{\mathcal{U}}_r$ and $\bm{\mathcal{V}}_r$, so the principal angles are small; in other words $\widetilde{\bm{\mathcal{U}}}_{r}$ and $\widetilde{\bm{\mathcal{V}}}_{r}$ are counted as good estimates for $\bm{\mathcal{U}}_r$ and $\bm{\mathcal{V}}_r$. Notice that we declare success when the normalized error $\leq 10^{-2}$ in $20$ iterations. We observe that GRMSPI needs much less measurements than ADMiRA to reach a fixed probability of success. 
\begin{figure*}
	\centering
	\mbox{\subfigure[]{\includegraphics[width=2.34in, height=1.5in]{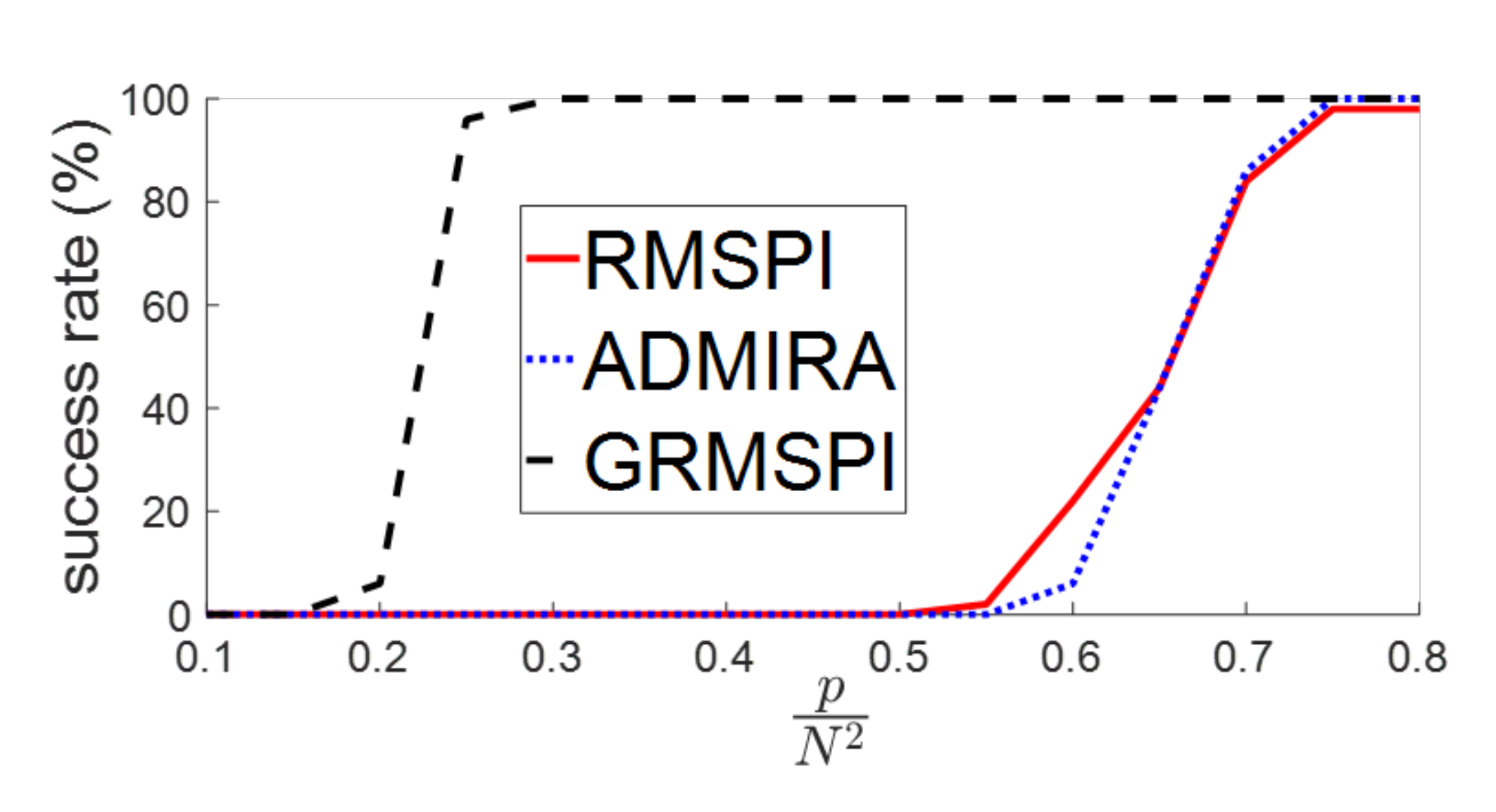}\label{fig2:a}}\quad
		\subfigure[]{\includegraphics[width=2.34in , height=1.5in]{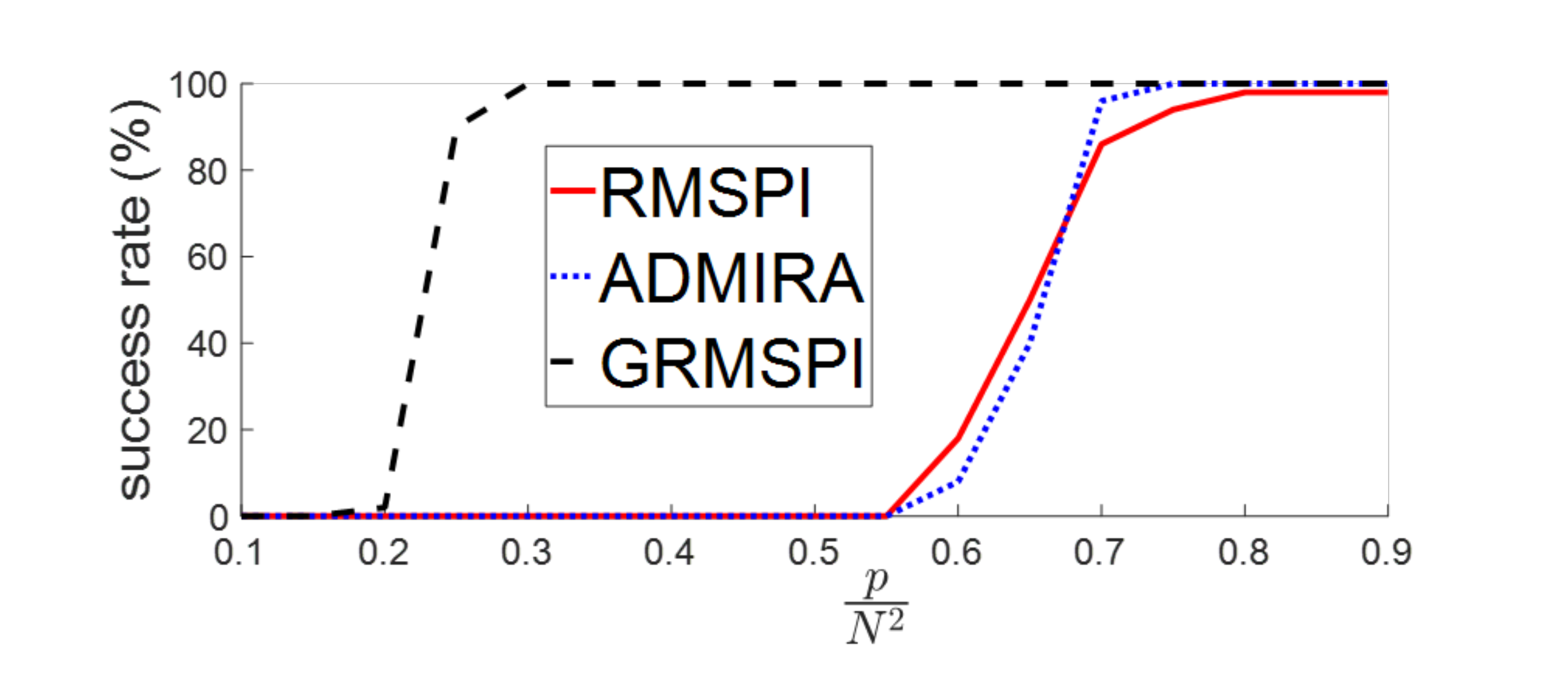}\label{fig2:b}}\quad
		\subfigure[]{\includegraphics[width=2.34in , height=1.5in]{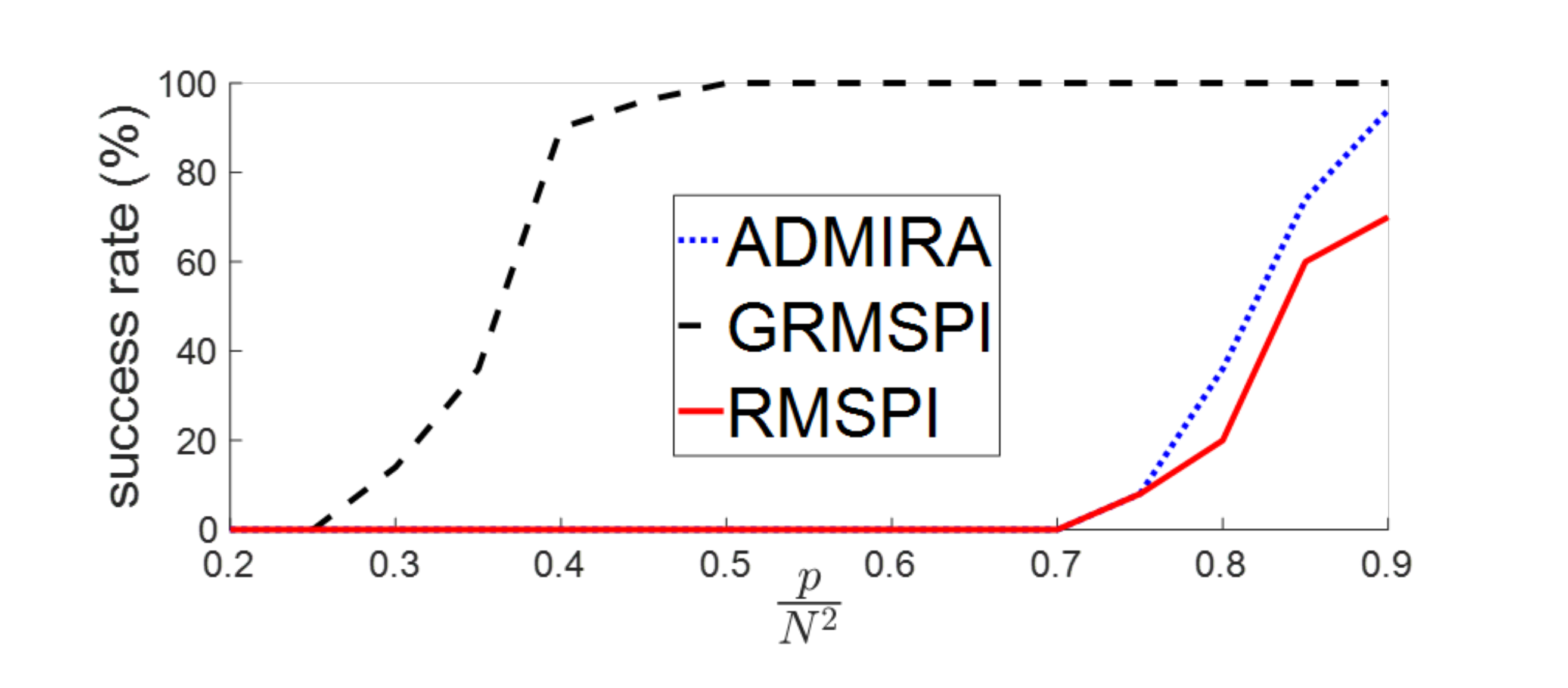}\label{fig2:c}}
	}
	\caption{Comparison of ADMiRA with RMSPI and GRMSPI in terms of success rate for (a) noiseless matrix recovery (b) noisy matrix recovery (c) matrix completion without noise. The respective principal angles are $\bm{\theta}_{u}=[89.8334,89.9545,89.9670]^{\rm{T}}$ and $\bm{\theta}_{v}=[89.7879,89.8493,89.9653]^{\rm{T}}$ and the weights are  $w_{2}=w_{4}=0.18$, $w_{1}=w_{3}=0.999$ ,$\bm{W}_{2}=\bm{W}_{4}=diag[0.17, 0.19,0.2]$, $ \bm{W}_{1}=diag[ 0.97,0.98,0.99] $ and $\bm{W}_{3}=diag[0.96, 0.97,0.99] $.}
	\label{fig2ADMIRAandprior0.05}
\end{figure*} 

Figure \ref{fig2:a},\ref{fig2:b}, \ref{fig2:c} show the success rate for noisy matrix recovery, noiseless matrix recovery and noiseless matrix completion, respectively. In this experiment we assume that $\widetilde{\bm{\mathcal{U}}}_{r}$ and $\widetilde{\bm{\mathcal{V}}}_{r}$ are close to $\bm{\mathcal{U}}_r^{\perp}$ and $\bm{\mathcal{V}}_r^{\perp}$ and the principal angles are small. We observe that the multi-weight scenario (GRMPSI) performs much better than the case of single weight (RMSPI).

\begin{figure*}
	\mbox{\subfigure[]{\includegraphics[width=2.34in, height=1.5in]{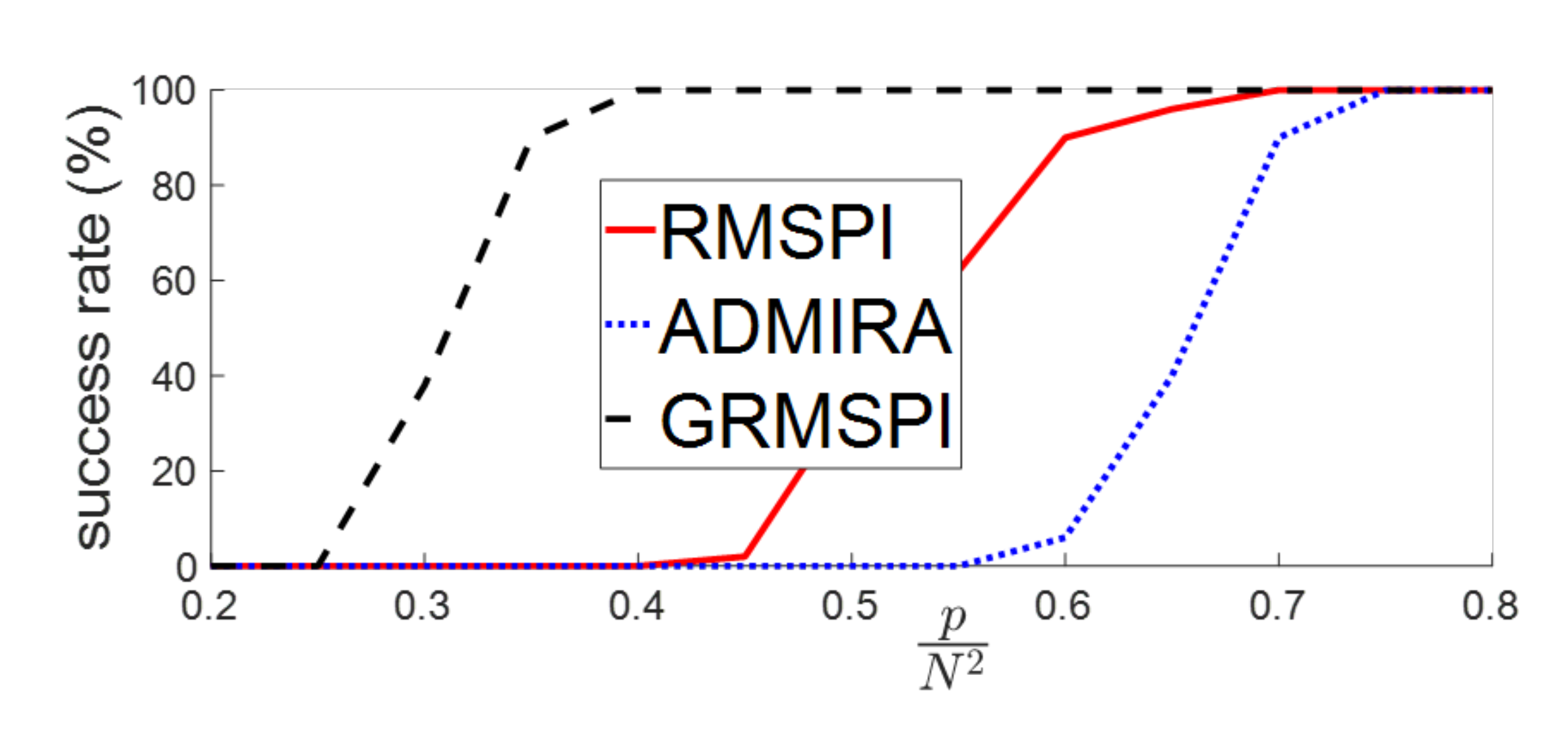}\label{fig3:a}}\quad
		\subfigure[]{\includegraphics[width=2.34in , height=1.5in]{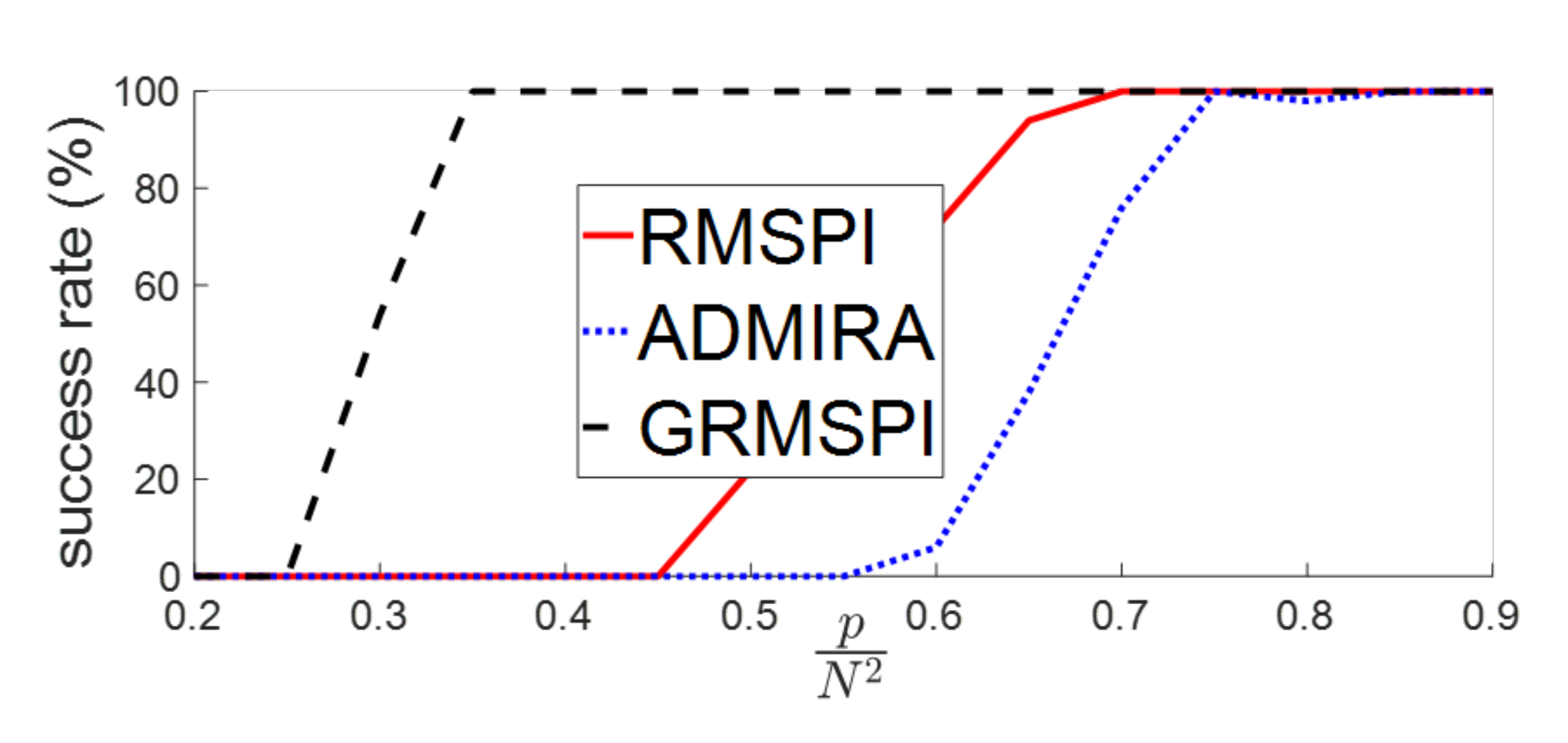}\label{fig3:b}}\quad
		\subfigure[]{\includegraphics[width=2.34in , height=1.5in]{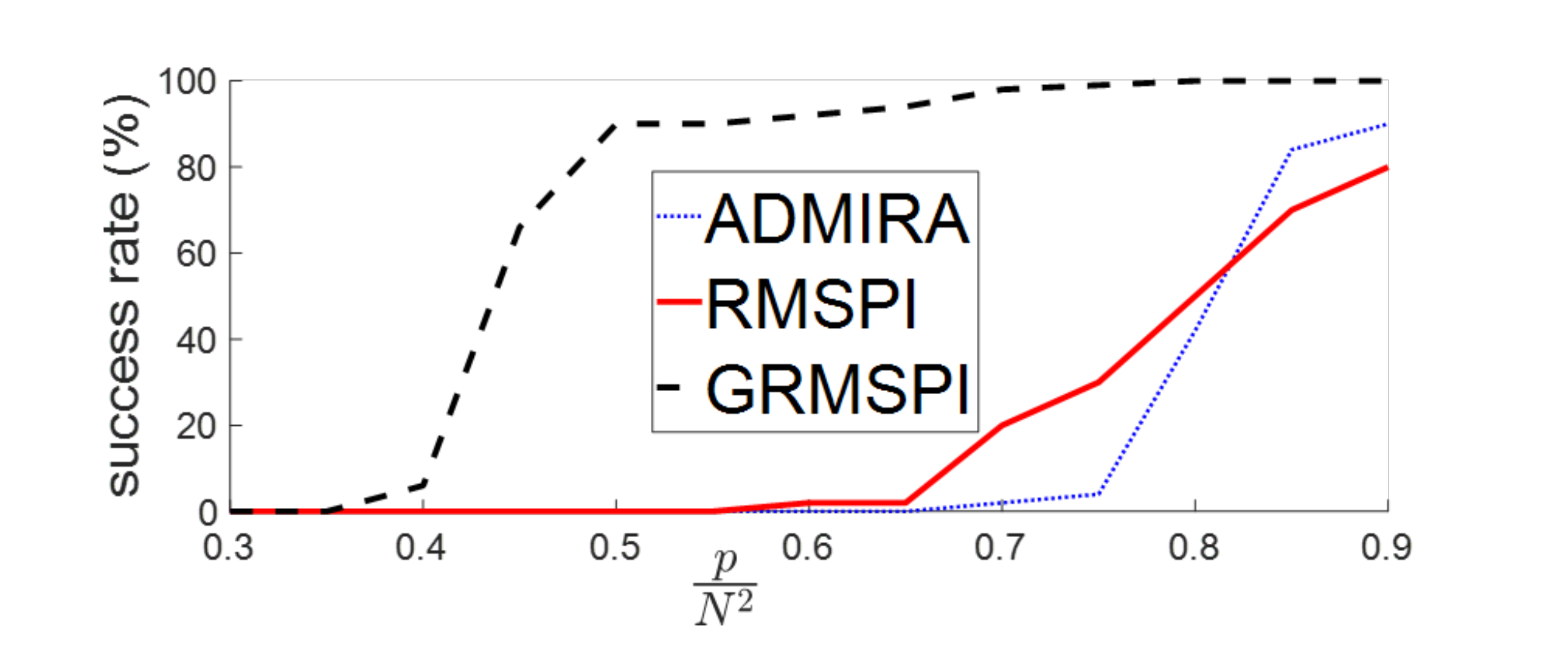}\label{fig3:c}}
	}
	\caption{Comparison of ADMiRA with RMSPI and GRMSPI in terms of success rate for (a) noiseless matrix recovery (b) noisy matrix recovery (c) matrix completion without noise. The respective principal angles are $\bm{\theta}_{u}=[2.5395,3.5460,3.6290]^{\rm{T}}$ and $\bm{\theta}_{v}=[89.8745,89.9585,89.9854]^{\rm{T}}$ and the weights are  $w_{1}=w_{4}=0.18$, $w_{2}=0.9556$,$w_{3}=0.999$ ,$\bm{W}_{1}=\bm{W}_{4}=diag[ 0.17, 0.19,0.21]$, $ \bm{W}_{2}=diag[ 0.93,0.94,0.95] $ and $\bm{W}_{3}=diag[0.97,0.98,0.99]$.}	\label{fig3ADMIRAandprior0.05}
\end{figure*}

\begin{figure*}
	\mbox{\subfigure[]{\includegraphics[width=2.34in, height=1.5in]{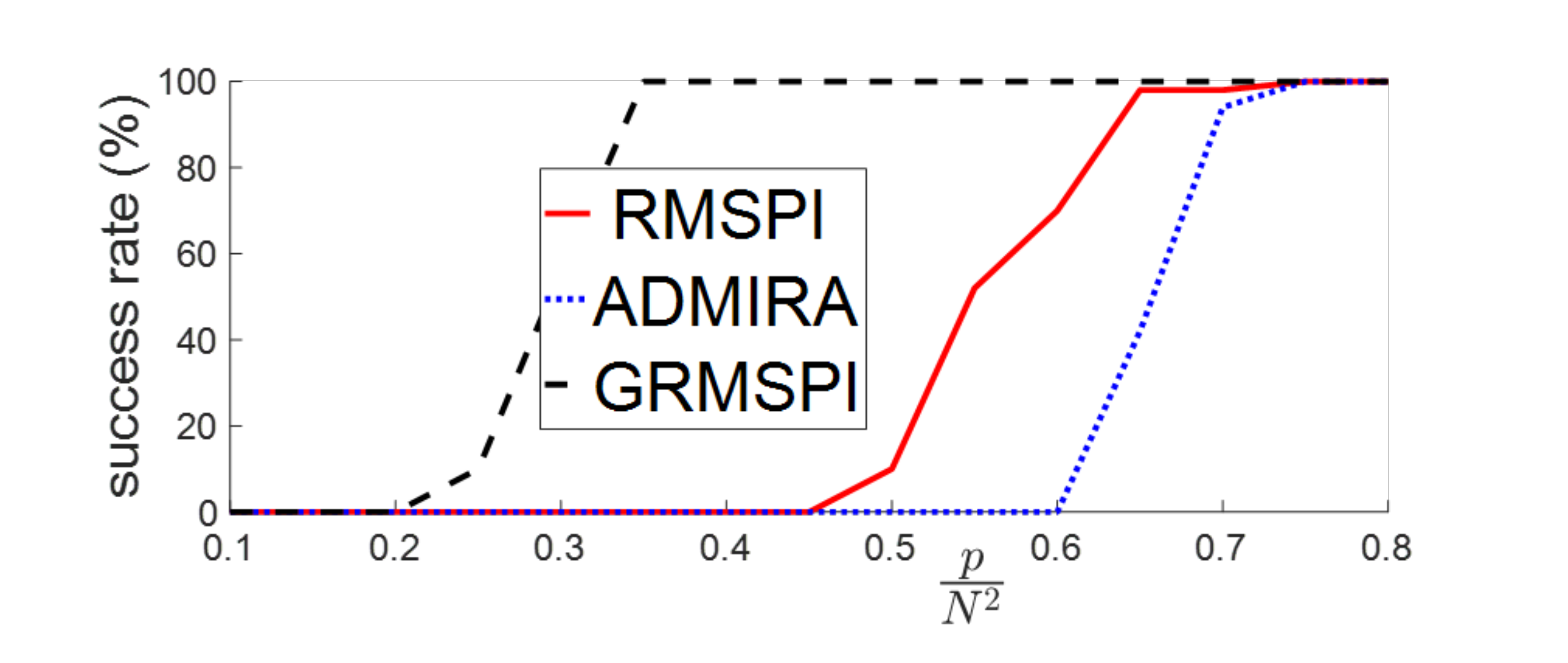}\label{fig4:a}}\quad
		\subfigure[]{\includegraphics[width=2.34in , height=1.5in]{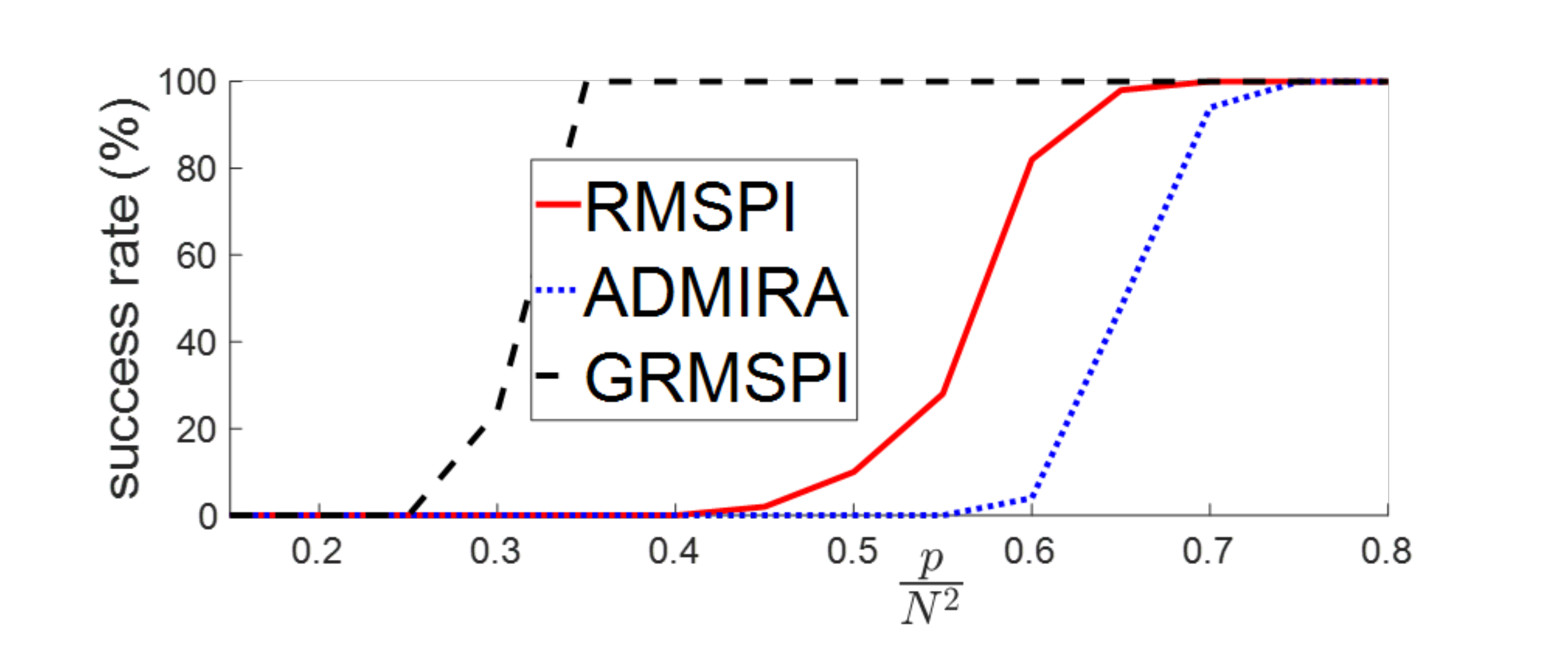}\label{fig4:b}}\quad
		\subfigure[]{\includegraphics[width=2.34in , height=1.5in]{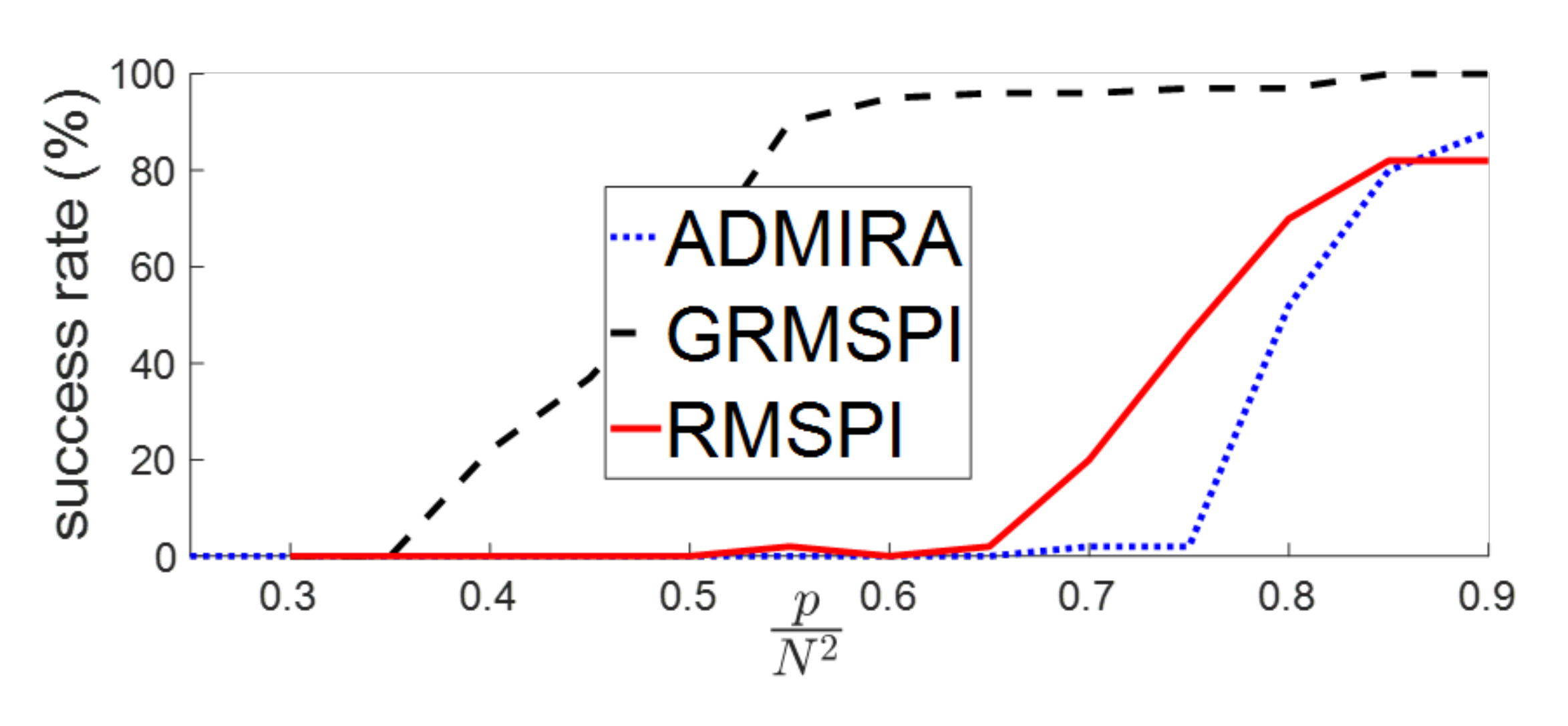}\label{fig4:c}}
	}
	\caption{Comparison of ADMiRA with RMSPI and GRMSPI in terms of success rate for (a) noiseless matrix recovery (b) noisy matrix recovery (c) matrix completion without noise. The respective principal angles are$\bm{\theta}_{u}=[89.8622,89.9070,89.9940]^{\rm{T}}$ and $\bm{\theta}_{v}=[2.4270,3.0595,3.6860]^{\rm{T}}$ and the weights are  $w_{2}=w_{3}=0.18$, $w_{1}=0.999$,$w_{4}=0.9576$ ,$\bm{W}_{2}=\bm{W}_{3}=diag[ 0.17, 0.19,0.21]$, $\bm{W}_{1}=diag[0.97,0.98,0.99]$ and $\bm{W}_{4}=diag[0.93, 0.94,0.95]$.}
	\label{fig4ADMIRAandprior0.05}
\end{figure*}

Figures \ref{fig3ADMIRAandprior0.05} and \ref{fig4ADMIRAandprior0.05} show the cases for which $\widetilde{\bm{\mathcal{U}}}_r$ and $\widetilde{\bm{\mathcal{V}}}_r$ are either close to or far from $\bm{\mathcal{U}}_r^{\perp}$ and $\bm{\mathcal{V}}_r ^{\perp}$, respectively. 
As expected, the performance of RMSPI and GRMSPI are better than the others. 

In Tables \ref{tabel1} to \ref{tabel4}, we provide experiments in which we compare the algorithms in terms of $ { \rm{SNR}} = 20\log(\frac{1}{{\rm normalized~ error}}) $, the number of required iterations for exact recovery in cases of different ranks. Notice that the considered cases for Tables \ref{tabel1} to \ref{tabel4} are respectively the same as in Figures \ref{fig1ADMIRAandprior0.05} to \ref{fig4ADMIRAandprior0.05}.
Notice that we do not consider ADMiRA in Tables \ref{tabel3}, \ref{tabel4}, since its performance is constant as it is not able to exploit the prior information. 

\begin{table*}
	\centering
	\begin{tabular}{|c|c|c|c|c|c|c|c|}
		\hline
		\multirow{2}{*}{$ r $} & \multirow{2}{*}{$ p/(m \times n) $} & \multicolumn{3}{c|}{$ SNR $} & %
		\multicolumn{3}{c|}{$ \sharp iter $}\\
		\cline{3-8}
		& & $ ADMiRA $  &$ RMSPI $ & $ GRMSPI $ & $ ADMiRA $ &  $ RMSPI $  & $ GRMSPI $  \\
		\hline\hline
		$ 2 $ &$  0.2  $& $ 12.66 $& $ 64.98 $ & $ 67.52$ & $ 29 $ & $ 9 $ & $ 11$\\ 
		\cline{2-8}
		&$  0.4  $& $ 50.54 $& $ 93.18 $ & $92.98 $ & $ 15 $ & $ 6 $ & $ 5$\\
		\cline{2-8}
		&$  0.6  $& $ 94.54 $& $ 95.70$  & $ 95.93$ & $ 11 $&$ 4 $ & $4 $\\
		\cline{2-8}
		&$  0.8  $& $ 96.68 $& $ 101.98 $ & $ 102.13$ & $ 7 $ & $ 4 $ & $ 4$\\
		\hline
		$ 5 $ &$  0.2  $& $ - $& $ 3.75 $& $4.42 $&$ - $&$ 5 $ &$5$ \\ 
		\cline{2-8}
		&$  0.4  $& $ 12.19 $& $ 7.25 $& $ 10.21 $&$ 11 $&$ 4 $ &$5$\\
		\cline{2-8}
		&$  0.6  $& $ 44.23 $& $ 18.58 $& $ 21.63$&$ 37 $&$ 5$ &$5$\\
		\cline{2-8}
		&$  0.8  $& $ 93.13 $& $ 42.55 $& $ 62.52 $&$ 23 $&$ 6 $&$6$\\
		\hline
		$ 10 $ &$  0.2  $& $1.77 $& $42.90$&$41.57$&$ 5 $&$ 6 $ &$6$\\ 
		\cline{2-8}
		&$  0.4  $& $5.23 $& $ 51.50 $&$51.12$&$ 7 $&$ 8 $&$7$\\
		\cline{2-8}
		&$  0.6  $& $ 12.52 $& $ 37.05 $&$38.35$&$ 7 $&$ 8 $&$7$\\
		\cline{2-8}
		&$  0.8  $& $ 30.23 $& $ 100.64 $&$101.34$&$ 9 $&$ 6 $&$5$\\
		\hline
	\end{tabular}
	\caption{Comparison of ADMiRA with RMSPI and GRMSPI in terms of SNR and number of required iterations when $\widetilde{\bm{\mathcal{U}}}_{r}$ and $\widetilde{\bm{\mathcal{V}}}_{r}$ are close to $\bm{\mathcal{U}}_{r}$ and $\bm{\mathcal{V}}_{r}$, respectively.}
	\label{tabel1}
\end{table*}
\begin{table*}
	\centering
	\begin{tabular}{|c|c|c|c|c|c|c|c|}
		\hline
		\multirow{2}{*}{$ r $} & \multirow{2}{*}{$ p/(m \times n) $} & \multicolumn{3}{c|}{$ SNR $} & %
		\multicolumn{3}{c|}{$ \sharp iter $}\\
		\cline{3-8}
		& & $ ADMiRA $  &$ RMSPI $ & $ GRMSPI $ & $ ADMiRA $ &  $ RMSPI $  & $ GRMSPI $  \\
		\hline\hline
		$ 2 $ &$  0.2  $& $ 13.09 $& $ 15.20 $ & $ 93.01 $ & $ 15 $ & $ 15 $ & $ 5 $\\ 
		\cline{2-8}
		&$  0.4  $& $ 47.99 $& $ 62.20 $ & $95.25 $ & $ 25 $ & $ 32 $ & $ 2$\\
		\cline{2-8}
		&$  0.6  $& $ 95.50 $& $ 94.20 $  & $ 101.10$ & $ 11 $&$ 11 $ & $2$\\
		\cline{2-8}
		&$  0.8  $& $ 92.33 $& $ 95.74 $ & $ 102.53 $ & $ 7 $ & $ 7 $ & $ 2$\\
		\hline
		$ 5 $ &$  0.2  $& $ 0.50 $& $ 1.67 $& $11.79 $&$ 5 $&$ 5 $ &$5$ \\ 
		\cline{2-8}
		&$  0.4  $& $ 13.24 $& $ 32.51 $& $ 89.96 $&$ 8 $&$ 28 $ &$10$\\
		\cline{2-8}
		&$  0.6  $& $ 45.21 $& $ 80.15 $& $ 96.75 $&$ 37 $&$ 33$ &$3$\\
		\cline{2-8}
		&$  0.8  $& $ 93.42 $& $ 88.88 $& $ 95.39 $&$ 23 $&$ 19 $&$2$\\
		\hline
		$ 10 $ &$  0.2  $& $1.80 $& $6.34$&$47.81$&$ 5 $&$ 5 $ &$6$\\ 
		\cline{2-8}
		&$  0.4  $& $5.19 $& $ 28.51 $&$64.26$&$ 7 $&$ 6 $&$7$\\
		\cline{2-8}
		&$  0.6  $& $ 12.36 $& $ 6464.66 $&$49.78$&$ 7 $&$ 7 $&$5$\\
		\cline{2-8}
		&$  0.8  $& $ 28.90 $& $ 94.56 $&$103.27$&$ 8 $&$ 6 $&$3$\\
		\hline
	\end{tabular}
	\caption{Comparison of ADMiRA with RMSPI and GRMSPI in terms of SNR and the number of required iteration when $\widetilde{\bm{\mathcal{U}}}_{r}$ and $\widetilde{\bm{\mathcal{V}}}_{r}$ are close to  $\bm{\mathcal{U}}_r^\perp$ and $\bm{\mathcal{V}}_r^{\perp}$, respectively.}
	\label{tabel2}
\end{table*}
\begin{table*}
	\centering
	\begin{tabular}{|c|c|c|c|c|c|}
		\hline
		\multirow{2}{*}{$ r $} & \multirow{2}{*}{$ p/(m \times n) $} & \multicolumn{2}{c|}{$ SNR $} & %
		\multicolumn{2}{c|}{$ \sharp iter $}\\
		\cline{3-6}
		&   &$ RMSPI $ & $ GRMSPI $  &  $ RMSPI $  & $GRMSPI $  \\
		\hline\hline
		$ 2 $ &$  0.2  $ & $ 15.86 $ & $ 88.55 $  & $ 9 $ & $ 12 $\\ 
		\cline{2-6}
		&$  0.4  $&  $ 91.12 $ & $97.06 $  & $ 18 $ & $ 4 $\\
		\cline{2-6}
		&$  0.6  $& $ 96.11 $  & $ 97.04$ &$ 7 $ & $3$\\
		\cline{2-6}
		&$  0.8  $&  $ 97.62 $ & $ 104.48$  & $ 5 $ & $ 3 $\\
		\hline
		$ 5 $ &$  0.2  $ & $ 0.98 $& $10.58 $&$ 5 $ &$5$ \\ 
		\cline{2-6}
		&$  0.4  $&  $ 7.87 $& $ 17.30 $&$ 4 $ &$5$\\
		\cline{2-6}
		&$  0.6  $& $ 76.91 $& $ 86.70$&$ 26 $ &$6$\\
		\cline{2-6}
		&$  0.8  $&  $ 91.97 $& $ 96.60 $&$ 14 $&$4$\\
		\hline
		$ 10 $ &$  0.2  $&  $19.24$&$44.59$&$ 5 $ &$6$\\ 
		\cline{2-6}
		&$  0.4  $&  $49.34 $&$56.94$&$ 6 $&$7$\\
		\cline{2-6}
		&$  0.6  $& $ 61.99 $&$46.99$&$ 8 $&$8$\\
		\cline{2-6}
		&$  0.8  $& $ 96.79 $&$101.42$&$ 4 $&$4$\\
		\hline
	\end{tabular}
	
	\caption{Comparison of ADMiRA with RMSPI and GRMSPI in terms of SNR and number of required iterations when  $\widetilde{\bm{\mathcal{U}}}_{r}$ and $\widetilde{\bm{\mathcal{V}}}_{r}$ are close to  $\bm{\mathcal{U}}_r$ and $\bm{\mathcal{V}}_r^{\perp}$, respectively.}
	\label{tabel3}
\end{table*}
\begin{table*}
	\centering
	\begin{tabular}{|c|c|c|c|c|c|}
		\hline
		\multirow{2}{*}{$ r $} & \multirow{2}{*}{$ p/(m \times n) $} & \multicolumn{2}{c|}{$ SNR $} & %
		\multicolumn{2}{c|}{$ \sharp iter $}\\
		\cline{3-6}
		&   &$ RMSPI $ & $ GRMSPI $  &  $ RMSPI $  & $ GRMSPI $  \\
		\hline\hline
		$ 2 $ &$  0.2  $ &$ 9.43 $ & $ 89.31 $  & $ 7 $ & $ 16 $\\ 
		\cline{2-6}
		&$  0.4  $&  $ 91.06 $ & $95.90$  & $ 15 $ & $ 4 $\\
		\cline{2-6}
		&$  0.6  $& $ 96.85 $  & $ 98.35$ &$ 7 $ & $3$\\
		\cline{2-6}
		&$  0.8  $&  $ 96.94 $ & $ 105.10$  & $ 5 $ & $ 3 $\\
		\hline
		$ 5 $ &$  0.2  $ & $ - $& $7.13 $&$ - $ &$5$ \\ 
		\cline{2-6}
		&$  0.4  $&  $ 8.20 $& $ 53.52 $&$ 4 $ &$9$\\
		\cline{2-6}
		&$  0.6  $& $ 78.22 $& $ 76.85$&$ 29 $ &$6$\\
		\cline{2-6}
		&$  0.8  $&  $ 91.30 $& $ 95.71 $&$ 15 $&$4$\\
		\hline
		$ 10 $ &$  0.2  $&  $118.82$&$44.00$&$ 5 $ &$6$\\ 
		\cline{2-6}
		&$  0.4  $&  $49.89 $&$55.10$&$ 6 $&$6$\\
		\cline{2-6}
		&$  0.6  $& $ 62.61 $&$46.82$&$ 8 $&$7$\\
		\cline{2-6}
		&$  0.8  $& $ 95.84 $&$103.08$&$ 4 $&$4$\\
		\hline
	\end{tabular}
	
	\caption{Comparison of ADMiRA with RMSPI and GRMSPI in terms of SNR and number of required iterations when $\widetilde{\bm{\mathcal{U}}}_{r}$ and $\widetilde{\bm{\mathcal{V}}}_{r}$ are close to  $\bm{\mathcal{U}}_r^{\perp}$ and $\bm{\mathcal{V}}_r$, respectively.}
	\label{tabel4}
\end{table*}
In these experiments, we observe that if the number of measurements increases, we have a better SNR and the number of required iterations decreases. Also, the performance will increase if the matrix of interest has a lower rank.
\begin{table*}
	\centering
	\begin{tabular}{|c|c|c|c|c|c|c|c|}
		\hline
		\multirow{2}{*}{$ r $} & \multirow{2}{*}{$ p/(m \times n) $} & \multicolumn{3}{c|}{$ SNR $} & %
		\multicolumn{3}{c|}{$ cpu time $}\\
		\cline{3-8}
		& & $ CVX $  &$ RMSPI $ & $ GRMSPI $ & $ CVX $ &  $ RMSPI $  & $ GRMSPI $  \\
		\hline\hline
		$ 2 $ &$  0.2  $& $ 442.41 $& $ 64.98 $ & $ 67.52$ & $ 15.42 $ & $ 0.37  $ & $ 0.30$\\ 
		\cline{2-8}
		&$  0.4  $& $ 392.75 $& $ 93.18 $ & $92.98 $ & $ 19.67 $ & $  0.28 $ & $0.26$\\
		\cline{2-8}
		&$  0.6  $& $420.72 $& $ 95.70$  & $ 95.93$ & $ 98.34 $&$  0.34 $ & $0.32 $\\
		\cline{2-8}
		&$  0.8  $& $452.33 $& $ 101.98 $ & $ 102.13$ & $ 44.05 $ & $  0.48$ & $ 0.44$\\
		\hline
		$ 5 $ &$  0.2  $& $ 58.61 $& $ 3.75 $& $4.42 $&$ 9.12 $&$ 0.23 $ &$0.21$ \\ 
		\cline{2-8}
		&$  0.4  $& $ 404.69 $& $ 7.25 $& $ 10.21 $&$ 19.52 $&$ 0.33 $ &$4.10$\\
		\cline{2-8}
		&$  0.6  $& $ 427.51 $& $ 18.58 $& $ 21.63$&$ 31.60 $&$ 0.54$ &$0.47$\\
		\cline{2-8}
		&$  0.8  $& $ 421.10 $& $ 42.55 $& $ 62.57 $&$ 197.37 $&$ 0.94 $&$0.82$\\
		\hline
		$ 10 $ &$  0.2  $& $36.38 $& $42.90$&$41.57$&$ 12.77 $&$ 0.51 $ &$0.47$\\ 
		\cline{2-8}
		&$  0.4  $& $62.42 $& $ 51.50$&$51.12$&$ 0.00 $&$ 1.42 $&$1.31$\\
		\cline{2-8}
		&$  0.6  $& $ 394.69 $& $ 37.05 $&$38.35$&$ 44.53 $&$ 2.59 $&$2.35$\\
		\cline{2-8}
		&$  0.8  $& $ 409.63 $& $ 100.64 $&$101.34$&$ 53.07 $&$ 2.48 $&$2.30$\\
		\hline
	\end{tabular}
	\caption{Comparison of CVX with RMSPI and GRMSPI in terms of SNR and cpu time when $\widetilde{\bm{\mathcal{U}}}_{r}$ and $\widetilde{\bm{\mathcal{V}}}_{r}$ are close to $ \bm{\mathcal{U}}_r $ and $ \bm{\mathcal{V}}_r $.
		}
	\label{tabel5}
\end{table*}

In Table \ref{tabel5}, we compare RMSPI, GRMSPI, and nuclear norm minimization implemented by CVX \cite{cvx} (succinctly shown by CVX) in terms of computational complexity and SNR. We observe that while CVX method outperforms the others in terms of SNR, its computational complexity is poorly large. Notice that \cite{eftekhari2018weighted} does not consider the case of far subspaces. Thus, it has not been included in this experiment. 
%

\appendices

\section{Proofs of least squares}\label{proof.least_square}
\begin{proof}	
	%
	To solve least squares in step 6 by using \eqref{p.6}, we solve the following least squares problem:
	\begin{align}
	\widetilde{\bm{X}} \leftarrow  \arg \underset{\bm{X}}{\min} \{\|\bm{y}-\mathcal{B}\bm{X}\|_2 : \bm{X} \in {\rm span}(\widetilde{\Psi})\}.
	\end{align}
	Due to the definition of $\mathcal{B}$ operator, the least squares problem is converted to
	\begin{align}
	\widetilde{\bm{x}} \leftarrow  \arg \underset{\bm{x}}{\min} \{\|\bm{y}-B\bm{x}\|_2\},
	\end{align}
	where
	\begin{align*}
	B = \begin{bmatrix}
	{\rm{vec}}(\bm{U}^{\rm{H}}\bm{Q}_{\widetilde{\bm{\mathcal{U}}}_{r}}^{-1}\bm{A}_{1}\bm{Q}_{\widetilde{\bm{\mathcal{V}}}_{r}}^{-1}\bm{V})^{\rm{T}}\\
	\vdots\\
	{ \rm{vec}}(\bm{U}^{\rm{H}}\bm{Q}_{\widetilde{\bm{\mathcal{U}}}_{r}}^{-1}\bm{A}_{i}\bm{Q}_{\widetilde{\bm{\mathcal{V}}}_{r}}^{-1}\bm{V})^{\rm{T}}  \\
	\vdots
	\\
	{\rm{vec}}(\bm{U}^{\rm{H}}\bm{Q}_{\widetilde{\bm{\mathcal{U}}}_{r}}^{-1}\bm{A}_{p}\bm{Q}_{\widetilde{\bm{\mathcal{V}}}_{r}}^{-1}\bm{V})^{\rm{T}} 
	\end{bmatrix} 	
	\end{align*}
	and 
	\begin{align*}
	\widetilde{\bm{x}} = B^{\dagger } \bm{y}.
	\end{align*}	
	Here, $ {\rm vec}$ is the operation that vectorizes a matrix by concatenation of the columns.		
	Finally, the solution of least squares reads
	\begin{align}
	\widetilde{\bm{X}} = \bm{U}~{\rm{reshape}}(\widetilde{\bm{x}})~\bm{V}^{\rm{H}},
	\end{align}	
	where ${\rm reshape}(\bm{x})$ write the vector $\bm{x}$ in a matrix form with size ...
\end{proof}
\section{Proof of Lemma \ref{lem.RIPrel}}\label{proof.lemRIP}
Suppose that $\delta(\mathcal{B})$ be the RIP constant of $\mathcal{B}$ i.e. we have
\begin{align}
(1-\delta_{r}(\mathcal{B}))\|\bm{Z}\|^{2}_{F} \leq \|\mathcal{B}\bm{Z}\|_{2}^{2} \leq (1+\delta_{r}(\mathcal{B}))\|\bm{Z}\|^{2}_{F}
\end{align}
for every $\bm{Z}$ with ${\rm rank}(\bm{Z})\le 1$ in particular $\bm{Z}'=\bm{Q}_{\widetilde{\bm{\mathcal{U}}}_{r}} \bm{Z} \bm{Q}_{\widetilde{\bm{\mathcal{V}}}_{r}}$ for any matrix $\bm{Z}$ with ${\rm rank}(\bm{Z})\le r$. Hence, we have:
\begin{align}\label{eq.relme1}
&\|\mathcal{B}(\bm{Z}')\|_2=\|\mathcal{A}(\bm{Z})\|_2\le (1+\delta(\mathcal{B}))\|\bm{Z}'\|_F\le\nonumber\\
& (1+\delta(\mathcal{B}))\|\bm{Q}_{\widetilde{\bm{\mathcal{U}}}_{r}}\|_{2\rightarrow 2}\|\bm{Z}\|_F\|\bm{Q}_{\widetilde{\bm{\mathcal{V}}}_{r}}\|_{2\rightarrow 2}\le (1+\delta(\mathcal{B}))\|\bm{Z}\|_F,
\end{align}
where in the first relation, we used the definition of $\mathcal{B}$. The last step uses the relations
\begin{align}
&\|\bm{Q}_{\widetilde{\bm{\mathcal{U}}}_{r}}\|_{2\rightarrow 2}\le 1\nonumber\\
&\|\bm{Q}_{\widetilde{\bm{\mathcal{V}}}_{r}}\|_{2\rightarrow 2}\le 1.
\end{align}
Since $\delta(\mathcal{A})$ is the smallest constant satisfying \eqref{eq.RIP_A}, it is straightforward from \eqref{eq.relme1} to verify that $\delta(\mathcal{A})\le \delta(\mathcal{B})$.
\section{Proof of convergence rate}\label{proof.conergancerate }
\begin{proof}
	In this section, we provide a new method for proving the convergence rate of our algorithm. Define $\bm{X}_1:=\bm{Q}_{\widetilde{\bm{\mathcal{U}}}_{r}}\bm{X}\bm{Q}_{\widetilde{\bm{\mathcal{V}}}_{r}}$. We should highlight that due to the invertiblity of $\bm{Q}_{\widetilde{\bm{\mathcal{U}}}_{r}}$ and $\bm{Q}_{\widetilde{\bm{\mathcal{V}}}_{r}}$, the convergence of $\widehat{\bm{X}}_{rec}$ to $\bm{X}$ is equivalent to the convergence of $\widehat{\bm{X}}$ to $\bm{X}_1$. Also, notice that the interested matrix $\bm{X}$ might not be exactly of rank $r$, hence we consider a rank-r truncation denoted by $\bm{X}_r$ and the residual $\bm{X}_r^{+}=\bm{X}-\bm{X}_r$. We begin with steps 7 and 8 of Algorithm \ref{algorithm.proposed}. We know that $\widehat{X}$ is a better estimate of $\bm{X}$ (in terms of $r$-term approximation) than
	 $\mathcal{P}_{\widetilde{\Psi}\cap {\rm supp}(\bm{X})}\bm{X}$. Due to the fact that $\widehat{\Psi} = {\rm supp}(\widehat{\bm{X}}^{k+1}) \subset \widetilde{\Psi} $, we conclude that
	

	\begin{align}
	& \|\mathcal{P}_{\widetilde{\Psi}}(\bm{X}_{1,r}-\widehat{\bm{X}}^{k+1})\|_{F} \leq \| \widetilde{\bm{X}} - \widehat{\bm{X}}^{k+1}\|_{F} +‌\| \widetilde{\bm{X}} - \mathcal{P}_{\widetilde{\Psi}}\bm{X}_{1,r}\|_{F} \nonumber\\
	& \leq 2\|\mathcal{P}_{\widetilde{\Psi}}(\bm{X}_{1,r} - \widetilde{\bm{X}}) \|_{F}‌.
	\end{align}
Considering that $\mathcal{P}_{\widetilde{\Psi}^{\bot}}\widehat{\bm{X}}^{k+1} = 0 $ and $ \mathcal{P}_{\widetilde{\Psi}^{\bot}}\widetilde{\bm{X}} = 0 $, we have 
	\begin{align}\label{eq.rel649}
	& \|\bm{X}_{1,r} - \widehat{\bm{X}}^{k+1}\|_{F}^{2} = \|‌\mathcal{P}_{\widetilde{\Psi}}(\bm{X}_{1,r} - \widehat{\bm{X}}^{k+1}) \|_{F}^{2} \\\nonumber 
	&+ \|‌\mathcal{P}_{\widetilde{\Psi}^{\bot}}(\bm{X}_{1,r} -\widetilde{\bm{X}}) \|_{F}^{2} \\ \nonumber
	&\leq 4\|\mathcal{P}_{\widetilde{\Psi}}(\bm{X}_{1,r} - \widetilde{\bm{X}}) \|_{F}^{2} + \|\mathcal{P}_{\widetilde{\Psi}^{\bot}}(\bm{X}_{1,r} - \widetilde{\bm{X}}) \|_{F}^{2}. 
	\end{align} 
	Let $ \widetilde{\bm{X}} $ be a solution of the following least squares problem (step 6 of Algorithm \ref{algorithm.proposed}).
	\begin{align*}
	\widetilde{\bm{X}} \leftarrow  \arg \underset{\bm{Z}}{\min} \{\|\bm{y}-\mathcal{B}\bm{Z}\|_2 :  \bm{Z} \in {\rm span}(\widetilde{\Psi})\}.
	\end{align*}
	Hence, by the properties of the least square solution, it holds that 
	\begin{align*}
	\langle \bm{y}-\mathcal{B}\widetilde{\bm{X}}, \mathcal{B}\bm{Z} \rangle =0  \quad  \forall \bm{Z}:~{\rm span}(\bm{Z})\subset \widetilde{\Psi}.
	\end{align*}
	In particular, we have $ \langle \mathcal{B}^{*}(\bm{y}-\mathcal{B}(\widetilde{\bm{X}})) ,\bm{Z} \rangle_{F} = 0 \quad  \forall \bm{Z}:~{\rm span}(\bm{Z})\subset \widetilde{\Psi} $ which is also equivalent to  $ \mathcal{P}_{\widetilde{\Psi}}\mathcal{B}^{*}(\bm{y}-\mathcal{B}(\widetilde{\bm{X}})) = 0  $. Due to $ \bm{y} = \mathcal{B}(\bm{X}_{1,r}) + \acute{\bm{e}} $ where $ \acute{\bm{e}} = \mathcal{B}(\bm{X}_{1,r}^{+})$, we have that 
	\begin{align*}
	\mathcal{P}_{\widetilde{\Psi}}\mathcal{B}^{*}(\mathcal{B}(\bm{X}_{1,r}-\widetilde{\bm{X}})) = -\mathcal{P}_{\widetilde{\Psi}}\mathcal{B}^{*}(\acute{\bm{e}}).
	\end{align*}
	By using the above equality, we may write
	\begin{align}\label{eq.rel11}
	&\|\mathcal{P}_{\widetilde{\Psi}}(\bm{X}_{1,r}-\widetilde{\bm{X}}) \|_{F} \nonumber\\
	& \leq \|\mathcal{P}_{\widetilde{\Psi}}((\mathcal{I} -\mathcal{B}^{*}\mathcal{B})(\bm{X}_{1,r}-\widetilde{\bm{X}})) \|_{F} +
	\| \mathcal{P}_{\widetilde{\Psi}}\mathcal{B}^{*}(\acute{\bm{e}}) \|_{F}‌  \nonumber\\
	& \leq \delta_{4r}(\mathcal{B}) \| \bm{X}_{r} - \widetilde{\bm{X}}\|_{F} +‌\| \mathcal{P}_{\widetilde{\Psi}}\mathcal{B}^{*}(\acute{\bm{e}}) \|_{F}‌,
	\end{align}
	where we used the relation
		\begin{align}\label{eq.rel616}
	&\|\mathcal{P}_{\widetilde{\Psi}}((\mathcal{I} -\mathcal{B}^{*}\mathcal{B})(\bm{X}_{1,r}-\widetilde{\bm{X}}))\|_F^2  	  \nonumber \\
	& = \langle \mathcal{P}_{\widetilde{\Psi}}((\mathcal{I} -\mathcal{B}^{*}\mathcal{B})(\bm{X}_{1,r}-\widetilde{\bm{X}})),(\mathcal{I} -\mathcal{B}^{*}\mathcal{B})\bm{X}\rangle _F \nonumber \\
	& \le \delta_{t} \|\mathcal{P}_{\widetilde{\Psi}}((\mathcal{I} -\mathcal{B}^{*}\mathcal{B})(\bm{X}_{1,r}-\widetilde{\bm{X}}))\|_F \|\bm{X}_{1,r}-\widetilde{\bm{X}}\|_F.
	\end{align}
	Here, $t\ge |\widetilde{\Psi} \cup {\rm supp}(\bm{X}_{1,r}-\widetilde{\bm{X}})|$.
	
	We proceed \eqref{eq.rel11} by writing
	\begin{align*}
	& \begin{bmatrix}
	\|\mathcal{P}_{\widetilde{\Psi}}\mathcal{B}^{*}(\mathcal{B}(\bm{X}_{1,r}-\widetilde{\bm{X}})) \|_{F} - ‌\| \mathcal{P}_{\widetilde{\Psi}}\mathcal{B}^{*}(\acute{\bm{e}}) \|_{F}
	\end{bmatrix}^{2} \\ \nonumber
	&\leq \delta_{4r}^{2}(\mathcal{B}) \| \bm{X}_{1,r} - \widetilde{\bm{X}}\|_{F}^{2} \nonumber\\
	&= \delta_{4r}^{2}(\mathcal{B})\Big(\|\mathcal{P}_{\widetilde{\Psi}}(\bm{X}_{1,r} - \widetilde{\bm{X}})\|_{F}^{2} + \|\mathcal{P}_{\widetilde{\Psi}^{\perp}}(\bm{X}_{1,r} - \widetilde{\bm{X}})\|_{F}^{2}\Big).
	\end{align*}
	Then, we use the equality $a^{2}-b^{2}=(a-b)(a+b)$ to reach  
	\begin{align}\label{eq.rel12}
	&\delta_{4r}^{2}(\mathcal{B}) \|\mathcal{P}_{\widetilde{\Psi}^{\perp}}(\bm{X}_{1,r} - \widetilde{\bm{X}})\|_{F}^{2} \geqslant (1-\delta_{4r}^{2}(\mathcal{B}))  \nonumber \\
	&\times\Big(\| \mathcal{P}_{\widetilde{\Psi}}(\bm{X}_{1,r} - \widetilde{\bm{X}})\|_{F} - \dfrac{1}{1+\delta_{4r}(\mathcal{B})}\|\mathcal{P}_{\widetilde{\Psi}}\mathcal{B}^{*}(\acute{\bm{e}})\|_{F}\Big)  \nonumber \\
	&\times\Big(\| \mathcal{P}_{\widetilde{\Psi}}(\bm{X}_{1,r} - \widetilde{\bm{X}})\|_{F} - \dfrac{1}{1-\delta_{4r}(\mathcal{B})}\|\mathcal{P}_{\widetilde{\Psi}}\mathcal{B}^{*}(\acute{\bm{e}})\|_{F}\Big).
	\end{align}
	By further simplifying \eqref{eq.rel12}, we have
	\begin{align*}
	&\dfrac{\delta_{4r}^{2}(\mathcal{B})}{(1-\delta_{4r}^{2}(\mathcal{B}))} \|\mathcal{P}_{\widetilde{\Psi}^{\perp}}(\bm{X}_{1,r} - \widetilde{\bm{X}})\|_{F}^{2}   \nonumber \\
	&\ge \Big(\| \mathcal{P}_{\widetilde{\Psi}}(\bm{X}_{1,r} - \widetilde{\bm{X}})\|_{F} - \dfrac{1}{1-\delta_{4r}(\mathcal{B})}\|\mathcal{P}_{\widetilde{\Psi}}\mathcal{B}^{*}(\acute{\bm{e}})\|_{F}^{2}\Big).
	\end{align*}
	By taking the square root, the above inequality reads
	\begin{align}
	& \|\mathcal{P}_{\widetilde{\Psi}}(\bm{X}_{1,r} - \widetilde{\bm{X}})\|_{F} \nonumber\\
	&\leq \dfrac{\delta_{4r}(\mathcal{B})}{\sqrt{1-\delta_{4r}^{2}(\mathcal{B})}} \|\mathcal{P}_{\widetilde{\Psi}^{\perp}}(\bm{X}_{1,r} - \widetilde{\bm{X}})\|_{F} \nonumber \\  
	&+ \dfrac{1}{1-\delta_{4r}(\mathcal{B})}\|\mathcal{P}_{\widetilde{\Psi}}\mathcal{B}^{*}(\acute{\bm{e}})\|_{F}.
	\end{align}
	Now, we intend to bound $\mathcal{P}_{\widetilde{\Psi}^{\perp}}(\bm{X}_{1,r}-\widetilde{\bm{X}})$. For this purpose, let $ \Phi={\rm{atom}}(\bm{X}_{1,r} - \widehat{\bm{X}}^{k}) $, then since $ |\Phi| \le {\rm rank}(\bm{X}_{1,r}) + {\rm rank}(\bm{X}^{k})\ \le 2r $, it holds that 
	\begin{align}\label{eq.rel13}
	& \|\mathcal{P}_{\Phi}\mathcal{B}^{*}(\bm{y}-\mathcal{B}(\widehat{\bm{X}}^{k}))\|_{F} \leq \|\mathcal{P}_{\acute{\Psi}}\mathcal{B}^{*}(\bm{y}-\mathcal{B}(\widehat{\bm{X}}^{k}))\|_{F}.
	\end{align}
 We can decompose each of the operators $ \mathcal{P}_{\Phi} $ and $ \mathcal{P}_{\acute{\Psi}} $ into two orthogonal projection operators as follow: 
 \begin{align*}
 \mathcal{P}_{\Phi}= \mathcal{P}_{\gamma} + \mathcal{P}_{\gamma^{\perp}} \mathcal{P}_{\Phi}, \\ 
\mathcal{P}_{\acute{\Psi}} =   \mathcal{P}_{\gamma} + \mathcal{P}_{\gamma^{\perp}} \mathcal{P}_{\acute{\Psi}},
 \end{align*}
 where $ {\rm span}(\gamma \subset \mathcal{O}) = {\rm span}(\Phi) \bigcap {\rm span}(\acute{\Psi})$. By applying this to \eqref{eq.rel13}, we reach 
\begin{align}\label{eq.rel14}
	& \|\mathcal{P}_{\gamma^{\perp}} \mathcal{P}_{\Phi}\mathcal{B}(\bm{y}-\mathcal{B}(\widehat{\bm{X}}^{k}))\|_{F} \leq \|\mathcal{P}_{\gamma^{\perp}} \mathcal{P}_{\acute{\Psi}}\mathcal{B}(\bm{y}-\mathcal{B}(\widehat{\bm{X}}^{k}))\|_{F}.
\end{align}
Since ${\rm supp}(\bm{X}_{1,r} - \widehat{\bm{X}}^{k})=\Phi$, the right-hand side is also equal to	
	
	\begin{align*}
	& \| \mathcal{P}_{\gamma^{\perp}}\mathcal{P}_{\acute{\Psi}}\mathcal{B}^{*}(\bm{y}-\mathcal{B}(\widehat{\bm{X}}^{k}))\|_{F}   \nonumber \\
	&\le \| \mathcal{P}_{\gamma^{\perp}}\mathcal{P}_{\acute{\Psi}}(\widehat{\bm{X}}^{k} - \bm{X}_{1,r}+\mathcal{B}^{*}(\bm{y}-\mathcal{B}(\widehat{\bm{X}}^{k})))\|_{F} \nonumber
	\end{align*}
	Also, the left-hand side can be bounded as 
	\begin{align*}
	\| &\mathcal{P}_{\gamma^{\perp}}\mathcal{P}_{\Phi}\mathcal{B}^{*}(\bm{y}-\mathcal{B}(\widehat{\bm{X}}^{k}))\|_{F} \geqslant  
	\|\mathcal{P}_{\acute{\Psi}^{\perp}}(\bm{X}_{1,r}-\widehat{\bm{X}}^{k})\|_{F}  \nonumber \\ & -\|\mathcal{P}_{\gamma^{\perp}}\mathcal{P}_{\Phi }(\widehat{\bm{X}}^{k}-\bm{X}_{1,r}+\mathcal{B}^{*}(\bm{y}-\mathcal{B}(\widehat{\bm{X}}^{k})))\|_{F}.
	\end{align*}
	Now the above inequalities imply that 
	\begin{align}\label{eq.rel15}
	&\| \mathcal{P}_{\acute{\Psi}^{\perp}}(\bm{X}_{1,r}-\widehat{\bm{X}}^{k})\|_{F}  \nonumber \\
	&\le \|\mathcal{P}_{\gamma^{\perp}}\mathcal{P}_{\Phi}(\widehat{\bm{X}}^{k}-\bm{X}_{1,r} +\mathcal{B}^{*}(\bm{y}-\mathcal{B}(\widehat{\bm{X}}^{k})))\|_{F} \nonumber \\
	&+\|\mathcal{P}_{\gamma^{\perp}}\mathcal{P}_{\acute{\Psi}}(\widehat{\bm{X}}^{k}-\bm{X}_{1,r}+\mathcal{B}^{*}(\bm{y}-\mathcal{B}(\widehat{\bm{X}}^{k})))\|_{F}  \nonumber \\
	&\leq \sqrt{2} \|\mathcal{P}_{\Phi\varDelta \acute{\Psi}}(\widehat{\bm{X}}^{k}-\bm{X}_{1,r}+\mathcal{B}^{*}(\bm{y}-\mathcal{B}(\widehat{\bm{X}}^{k})))\|_{F} \nonumber \\
	& \leq \sqrt2{} \|\mathcal{P}_{\Phi\varDelta \acute{\Psi}}((\mathcal{I}-\mathcal{B}^{*}\mathcal{B})(\widehat{\bm{X}}^{k} - \bm{X}_{1,r}))\|_{F} \nonumber\\
	&+ \sqrt{2} \|\mathcal{P}_{\Phi\varDelta \acute{\Psi}}\mathcal{B}^{*}(\acute{\bm{e}})\|_{F}.
	\end{align}
	In \eqref{eq.rel15}, $ \Phi\varDelta \acute{\Psi} $ is the symmetric difference of the sets $ \Phi$ and $ \acute{\Psi} $. Since $\acute{\Psi}\subseteq \widetilde{\Psi}$ by step 5, thus $\widetilde{\Psi}^{\perp} \subseteq\acute{\Psi}^{\perp}$. Consequently, we have 
	\begin{align*}
	&\|\mathcal{P}_{\acute{\Psi}^{\perp}}(\bm{X}_{1,r}-\widehat{\bm{X}}^{k})\|_{F} \geqslant \|\mathcal{P}_{\widetilde{\Psi}^{\perp}}\bm{X}_{1,r}-\widehat{\bm{X}}^{k})\|_{F}  \nonumber \\
	& =\|\mathcal{P}_{\widetilde{\Psi}^{\perp}}\bm{X}_{1,r}\|_{F} = \|\mathcal{P}_{\widetilde{\Psi}^{\perp}}(\bm{X}_{1,r}-\widetilde{\bm{X}})\|_{F},
	\end{align*}
	where we used the facts that $\mathcal{P}_{\widetilde{\Psi}^{\perp}}\widehat{\bm{X}}^{k}$ and $\mathcal{P}_{\widetilde{\Psi}^{\perp}}\widetilde{\bm{X}}$. By considering the above bound and \eqref{eq.rel616}, we obtain
	\begin{align}\label{eq.rel651}
	&\|\mathcal{P}_{\widetilde{\Psi}^{\perp}}(\bm{X}_{1,r}-\widetilde{\bm{X}})\|_{F} \leq \sqrt{2}\delta_{4r}(\mathcal{B})\|\widehat{\bm{X}}^{k}-\bm{X}_{1,r}\|_{F} + \nonumber\\
	&\sqrt{2} \|\mathcal{P}_{\Phi\varDelta \acute{\Psi}}\mathcal{B}^{*}(\acute{\bm{e}})\|_{F}.
	\end{align}
	Now, we aim at bounding $\|\bm{X}_{1,r} - \widehat{\bm{X}}^{k+1}\|_{F}$ in \ref{eq.rel649} by combining \eqref{eq.rel15}, \eqref{eq.rel651}. This leads to
	\begin{align}\label{eq.relfinal}
	&\|(\bm{X}_{1,r}- \widehat{\bm{X}}^{k+1})\|_{F}^{2} \leq \|\mathcal{P}_{\widetilde{\Psi}^{\perp}}(\bm{X}_{r}-\widetilde{\bm{X}})\|_{F}^{2}  \\
	&+ 4(\dfrac{\delta_{4r}(\mathcal{B})}{\sqrt{1-\delta_{4r}^{2}(\mathcal{B})}} \|\mathcal{P}_{\widetilde{\Psi}^{\perp}}(\bm{X}_{1,r} - \widetilde{\bm{X}})\|_{F}  \\
	&+ \dfrac{1}{1-\delta_{4r}(\mathcal{B})}\|\mathcal{P}_{\widetilde{\Psi}}\mathcal{B}^{*}(\acute{\bm{e}})\|_{F})^{2}  \\
	&\leq (\sqrt{\dfrac{1+3\delta_{4r}^{2}(\mathcal{B})}{1-\delta_{4r}^{2}(\mathcal{B})}} \|\mathcal{P}_{\widetilde{\Psi}^{\perp}}(\bm{X}_{1,r} - \widetilde{\bm{X}})\|_{F}  \\
	&+ \dfrac{1}{1-\delta_{4r}(\mathcal{B})}\|\mathcal{P}_{\widetilde{\Psi}}\mathcal{B}^{*}(\acute{\bm{e}})\|_{F})^{2} 
	\end{align}
	where we used the relation $a^{2}+(b+c)^{2} \leq (\sqrt{a^{2}+b^{2}}+c)^{2}$ for the last inequality. The only point that remains is to replace \eqref{eq.rel651} into \eqref{eq.relfinal} as follows.
	\begin{align*}
	&\|(\bm{X}_{1,r}- \widehat{\bm{X}}^{k+1})\|_{F}  \nonumber \\
	&\leq \sqrt{\dfrac{2\delta_{4r}^{2}(\mathcal{B})(1+3\delta_{4r}^{2}(\mathcal{B}))}{1-\delta_{4r}^{2}(\mathcal{B})}}\|(\bm{X}_{1,r}- \widehat{\bm{X}}^{k})\|_{F}  \nonumber \\ 
	&+ \sqrt{\dfrac{2(1+3\delta_{4r}^{2}(\mathcal{B}))}{1-\delta_{4r}^{2}(\mathcal{B})}} \|\mathcal{P}_{\Phi\varDelta \acute{\Psi}}\mathcal{B}^{*}(\acute{\bm{e}})\|_{F} \nonumber \\
	&+ \dfrac{2}{1-\delta_{4r}(\mathcal{B})} \|\mathcal{P}_{\widetilde{\Psi}}\mathcal{B}(\acute{\bm{e}})\|_{F}.
	\end{align*}
	For convergence, we must have $$\rho := \sqrt{\dfrac{2\delta_{4r}^{2}(\mathcal{B})(1+3\delta_{4r}^{2}(\mathcal{B}))}{1-\delta_{4r}^{2}(\mathcal{B})}} \le 1.$$	
It is straightforward to verify that this requires $ 6\delta_{4r}^{4}+ 3\delta_{4r}^{2} - 1 < 0 $ which could be further simplified to 
	\begin{align*}
	\delta_{4r}^{2}(\mathcal{B}) < \dfrac{\sqrt{\frac{11}{3}}-1}{4} \nonumber \\
	\delta_{4r}(\mathcal{B}) < 0.4782.
	\end{align*}
\end{proof} 

\ifCLASSOPTIONcaptionsoff
\newpage
\fi
\bibliographystyle{ieeetr}
\bibliography{Myrefrence}
\end{document}